\documentclass[fleqn,usenatbib]{mnras}

\usepackage{newtxtext,newtxmath}
\usepackage[T1]{fontenc}
\usepackage{ae,aecompl}

\usepackage{graphicx}	
\usepackage{amsmath}	
\usepackage{amssymb}	
\usepackage{subfigure}

\usepackage[table,xcdraw]{xcolor}

\title[Asymmetric mid-plane gas in HD~100546]{Asymmetric mid-plane gas in ALMA images of HD~100546}

\author[J. M. Miley et al.]{J. M. Miley$^{1}$\thanks{E-mail: py12jm@leeds.ac.uk},
O. Pani\'c $^{1}$,
T. J. Haworth $^{2}$,
I. Pascucci $^{3}$,
M. Wyatt $^{4}$,
C. Clarke  $^{4}$, 
\newauthor
A. M. S. Richards $^{5}$,
T. Ratzka $^{6}$ 
\\
$^{1}$School of Physics \& Astronomy, University of Leeds, Woodhouse Lane, Leeds, LS2 9JT, UK \\
$^{2}$Astrophysics Group, Imperial College London, Blackett Laboratory, Prince Consort Road, London SW7 2AZ, UK\\
$^{3}$Lunar and Planetary Laboratory, The University of Arizona, Tucson, AZ 85721, USA\\
$^{4}$Institute of Astronomy, Madingley Rd, Cambridge, CB3 0HA, UK\\
$^{5}$Jodrell Bank Centre for Astrophysics, School of Physics and Astronomy, University of Manchester, Manchester M13 9PL, UK\\
$^{6}$Universitaets-Sternwarte Muenchen, Ludwig-Maximilians-Universitaet, Scheinerstr. 1, 81679 Muenchen, Germany \\
}

\date{Accepted XXX. Received YYY; in original form ZZZ}

\pubyear{2018}

\begin{document}
\label{firstpage}
\maketitle

\begin{abstract}
In this paper we present new ALMA observations towards the proto-planet hosting transitional disc of Herbig Ae/Be star HD 100546. This includes resolved 1.3~mm continuum, $^{13}$CO and the first detection of C$^{18}$O in this disc, which displays azimuthal asymmetry in regions spatially coincident with structures previously identified in HST images related to spiral arms. 
The lower limit on the mass of the dust disc is calculated to be 9.6~$\times 10^{-4} \rm ~$M$_\odot$. A firm lower-limit on the total gas mass calculated from optically thin, mid-plane tracing C$^{18}$O (2-1) emission is 0.018 M$_\odot$ assuming ISM abundances. These mass estimates provide an estimate of gas-to-dust ratio in the disc of 19, the ratio will increase if C$^{18}$O is relatively under-abundant in the disc compared to CO and H$_2$. 
Through deprojection and azimuthal averaging of the image plane we detect 1.3~mm continuum emission out to 290$\pm$10~au, $^{13}$CO to 390$\pm$10~au and C$^{18}$O to 300$\pm10$~au. 
We measure a radially increasing millimetre spectral index between wavelengths of 867~$\mu$m and 1.3~mm, which shows that grain sizes increase towards the star, with solid particles growing to cm scales in the inner disc.

\end{abstract}

\begin{keywords}
protoplanetary discs -- techniques: interferometric -- stars: pre-main sequence
\end{keywords}


\section{Introduction}
\label{sec:intro}
Interferometric observations at millimetre and sub-millimetre wavelengths often reveal millimetre-sized particles in the outer regions of Herbig and T-Tauri discs. Radial drift timescales for particles in the outer regions of the disc ($>$100~au) are of order $10^5$years, however disc lifetimes are of order $10^6-10^7$ years, meaning the particles in the outer disc should have already drifted onto the star. Retention of large grains in the outer disc, or alternatively some form of continuous replenishment, therefore seems to be essential in order to explain the observational data \citep{Klahr2006,Birnstiel2013}. 
The consequences of mechanisms that concentrate disc material locally are now becoming frequently observed. Results from ALMA have revealed highly asymmetric, so called `lop-sided discs' such as SAO 206462 and SR 21 \citep{Perez2014}, IRS 48 \citep{vanderMarel2015} and HD 142527 \citep{Casassus2015}. Radial gas pressure bumps can also result in dust traps that form symmetric, concentric rings in the dust disc \citep{Pinilla2012RingDisks,Dullemond2018}.


Continuum emission at mm-wavelengths is also a diagnostic tool for tracing the dust in a protoplanetary disc, specifically the coolest dust contained in the mid-plane and in the outer disc, where the majority of solid mass is expected to be contained.
Measured flux can be related to the mass of emitting particles by assuming optically thin emission and using
\begin{equation}
\label{eqn:dust_mass}
F_\nu = \frac{\kappa_\nu B_\nu(T) M_{dust}}{D^{2}},
\end{equation}
where F$_\nu$ is the continuum flux measured at frequency $\nu$, $\kappa_\nu$ is the specific opacity and D is the distance to the source from the observer. 
Continuum emission also depends on particle size; millimetre continuum emission comes most efficiently from a narrow range of grain sizes between 1mm and 1cm \citep{Takeuchi2005}. If grains have grown beyond $ a_{max} \approx 3\lambda$, where $\lambda$ is the wavelength of observation, they emit less efficiently \citep{Draine2006}. 
Radial drift and grain growth predict the largest solid particles to be concentrated toward the star, leaving the smaller ones behind. As emission brightness is related to grain size we can therefore expect a variation in observed disc size at different observing wavelengths, but also a variation in spectral index, $\alpha_{\rm mm}$, across the disc. Spatially resolved continuum images of discs at multiple wavelengths allow for a calculation of a position-dependent $\alpha_{\rm mm}$, where, for optically thin emission, S$_\nu \propto \nu^{\alpha_{\rm mm}}$ (e.g. HD 163296, \citealt{Guidi2016}). Surveys of star forming regions find that disc-averaged spectral indices can lie in the range 1.5 $\le \alpha_{\rm mm} \le$ 3.5  \citep{Ricci2012,Pinilla2014}. If the size of grains in the disc varies with radius, then so will the spectral index.

\citet{Draine2006} computed the frequency dependent opacity of different grain size distributions. A calculation of the opacity index $\beta$, where $\kappa_\nu \propto \nu^{\beta}$, between two spatially resolved observations at different frequency can therefore provide a quantitative estimate as to how large grains have grown in different parts of the disc by a comparison to the calculations of \citet{Draine2006}. $\beta$ values evolve with grain growth; $\beta \approx$ 1.5-2 is associated with the ISM \citep{Mathis1977}, and for the vast majority of disc masses, these `lower' values of $\beta \le 1$ are likely to be as a result of grain growth rather than other potential explanations such as optically thick emission \citep{Ricci2012}.   

In order to calculate a reliable total mass of the disc we must have a robust gas mass measurement \citep{Williams2014} as well as one for the dust. The mass of the dust disc is canonically related to the gas mass by an empirical ratio taken from the ISM; g/d (gas-to-dust ratio) = 100. Modelling of gas emission has increasingly suggested this may not be the case however.
A complication is that modelling of the gaseous spectral lines is generally degenerately sensitive to g/d - that there is no unique g/d that satisfies the observations \citep{Bruderer2012,Boneberg2016}. 
Using ALMA data, a recent survey of the Lupus star forming region gives evidence that supports low observed g/d in protoplanetary discs; \citet{Miotello2017} find that 23 out of their 34 discs with gas and dust detections have g/d < 10. Chemical reprocessing can leave CO depleted in discs, with CO locked into larger icy bodies \citep{Du2015} or converted into complex molecules \citep{Kama2016,Yu2016}. The carbon chemistry that controls the abundances is highly sensitive to the temperature profiles of the dust within the disc \citep{Reboussin2015}.


A complication for inferring gas properties such as the total mass is that gas emission is typically either optically thick or requires high sensitivity to detect. In this study we observe C$^{18}$O, a less optically thick alternative to other CO isotopologues \citep{VanZadelhoff2001,Dartois2003}. The trade-off here is that high sensitivity is required to detect a molecule much less abundant than other gas tracers. 
For example in the ALMA survey of Lupus, \citet{Ansdell2016} detect C$^{18}$O (3-2) emission in 11 out of their 89 targets with integrations of between 30-60 seconds per source. Detection rates are higher for intermediate mass Herbig stars, where the brighter star supports a warmer disk meaning more C$^{18}$O is found in the gas phase.
To date there is only a handful of C$^{18}$O detections in the discs of Herbig stars, including MWC 480 \citep{Guilloteau2013}, HD 169142 \citep{Panic2008,Fedele2017}, HD 163296 \citep{Rosenfeld2013} and AB Aur \citep{Pacheco-Vazquez2016}. When the optically thin C$^{18}$O and other isotopologues are observed it permits us to compute the CO disc mass and temperature. Emission from C$^{18}$O traces the dense disc mid-plane \citep{VanZadelhoff2001,Dartois2003}, where we expect grain growth and the formation of giant planets. Resolved detections of emission from this molecule will therefore allow us to study a part of the disc that is crucial for planet formation.

Much recent work has focused on transition discs. Transition discs are a class of protoplanetary discs that have a lesser flux in the mid-IR due to the opening of a gap or cavity in the inner disc which is generally explained by some combination of mechanisms such as grain-growth \citep{Tanaka2005}, photo evaporation \citep{Owen2016,Ercolano2017TheObservations}, clearing by a forming exoplanet or binary companion \citep{Baruteau2014,Price2018} and dead-zones \citep{Dzyurkevich2010TrappingSimulations}. This clearing eventually leads to gas-poor, optically thin, debris discs \citep{Espaillat2014,Wyatt2015}. There is now an increasing number of observational examples of resolved inner cavities of transition discs \citep{Williams2011,vanderMarel2017,Pinilla2018HomogeneousCavities}.  Characterising the distribution of gas and dust in such systems may be able to shed light on how this transition occurs. 

In this paper we present new ALMA observations towards Herbig Ae/Be star HD 100546 in order to characterise the dust and gas in its disc. In the next section we briefly review recent and related published work on HD 100546. 
In Section \ref{sec:obs} we describe the data taken and then present the initial results and measurements in Section \ref{sec:results}. In Section \ref{sec:AD_mass} we calculate the gas and dust mass in the disc and assess apparent asymmetry in the interferometric images produced. We use the deprojected, azimuthally averaged flux profiles to determine the radial extent of the emission for each of the observed tracers and calculate $\beta_{\rm mm}$ to infer levels of grain growth. Our main results are summarised in Section \ref{sec:conc}.

\section{HD 100546}
\label{sec:hd100}
This study focuses on a fascinating example of planet formation; the well-studied disc around Herbig Ae/Be star HD 100546. The star is located at a distance of 110pc \citep{Lindegren2018} with a mass reported by \citet{vandenAncker1997} of 2.4 M$_{\odot}$ from pre-main sequence (PMS) track fitting and more recently 1.9 M$_{\odot}$ by \citet{Fairlamb2015} derived through X-shooter spectroscopy and PARSEC PMS tracks \citep{Bressan2012}. Although previously thought to be as old as  $\ge$10Myr \citep{vandenAncker1997}, \citet{Fairlamb2015} calculate an age for the star of 7.02$\pm$1.49 Myr and using the updated parallax from GAIA data release 2, \citet{Vioque2018} find a younger age of 5.5$^{+1.4}_{-0.7}$ Myr.

\subsection{Evidence for proto-planet(s)} 
The system has been long associated with active planet formation and so is well studied. A protoplanet has been confirmed at $\sim$50~au via direct imaging 
\citep{Quanz2015CONFIRMATIONAU} 
and there has been some evidence of an inner companion as well. This includes SED modelling and mid-IR interferometry \citep{Mulders2013,Panic2014}
, ALMA observations 
\citep{Walsh2014}
, near IR spectroscopy 
\citep{Brittain2013,Brittain2014} 
and a point source in GEMINI Planet Imager (GPI) data at 10~au \citep{Currie2015}
support the potential HD 100546`c'. However this predicted companion has alluded detection in more recent GPI and Magellan Adaptive Optics System observations by \citet{Follette2017}. The position of the potential inner companion coincides with the position of the rim of the inner gap at $\sim11\pm1$~au, interior to which is a small inner dust disc of $\le$0.7~au \citep{Panic2014} and a recently detected bar-like structure reaching across the inner gap in both continuum and H$\alpha$ emission \citep{Mendigutia2017b}.

\subsection{Structure in the disc} 
Observing with ATCA, 7mm continuum emission can be fit by a Gaussian with FWHM between 50 and 60 au \citep{Wright2015}, and the authors assert that grains in the disc are well-processed and have grown as large as 5cm and possibly further. 
\citet{Walsh2014} are able to detect a significant amount of weak, extended continuum emission beyond this and find a best-fit model for the dust in the uv plane comprised of two rings with a Gaussian profile; one centred at 26~au with a FWHM of 21~au and one at 190~au with FWHM 75~au. \citet{Pineda2014} find that mm-sized grains are trapped within a similar inner ring and note brightness asymmetries in the dust between the South-East and North-West of the disc. 
All three of these studies at (sub-)millimetre wavelengths favour arguments of at least one giant planet orbiting within the disc. The disc does not appear to be axi-symmetric; scattered light studies routinely observe multiple asymmetric features and spiral arms within the dust disc \citep{Garufi2016,Follette2017}, including a `dark lane' observed by \citet{Avenhaus2014} in the same direction as the horse-shoe shaped asymmetry identified in 7~mm images of \citep{Wright2015}. The exact nature of this asymmetry at millimetre wavelengths has until now remained largely undefined. 

The gas in the disc is much more spatially extended than the dust, CO has been measured out to $\sim$400au \citep{Walsh2014}. Very little is known about the gas mass of this uniquely interesting disc due to a lack of data observing optically thin gas lines. Gas mass calculations based on optically thick $^{12}$CO emission can lead to underestimations of as large as two orders of magnitude in comparison to those using optically thin CO isotopologues \citep{Moor2011,Miley2018}.

\begin{figure*}
\includegraphics[width=\linewidth]{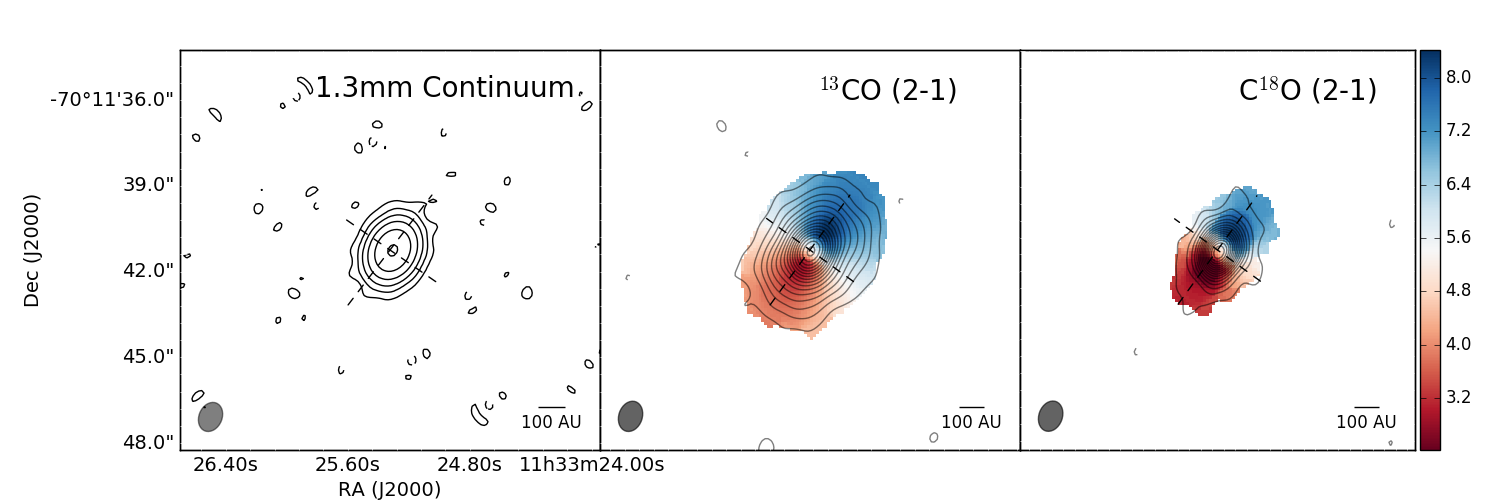}
\caption{ On the left is the contour map of the continuum emission, contours increase in base 3 logarithms from 3$\sigma$ to 729$\sigma$. Intensity weighted velocity maps of the two isotopologue transitions are shown overlaid with contours of the integrated line emission. For $^{13}$CO, contours start at 3$\sigma$ and rise in steps of 5$\sigma$ to 78$\sigma$, for C$^{18}$O(2-1) contours start at 3$\sigma$, and rise in steps of 3$\sigma$ up to 45$\sigma$. Dashed contours are negative. Beam ellipses and a scale bar of 100 au are shown in each image. North is up and West is right, dashed lines follow the major and minor axis of the disc, assuming a position angle of 144$^{\circ}$ and join at the stellar position given by GAIA.}
\label{fig:ims}
\end{figure*}

\begin{figure}
\centering \includegraphics[width=1\linewidth]{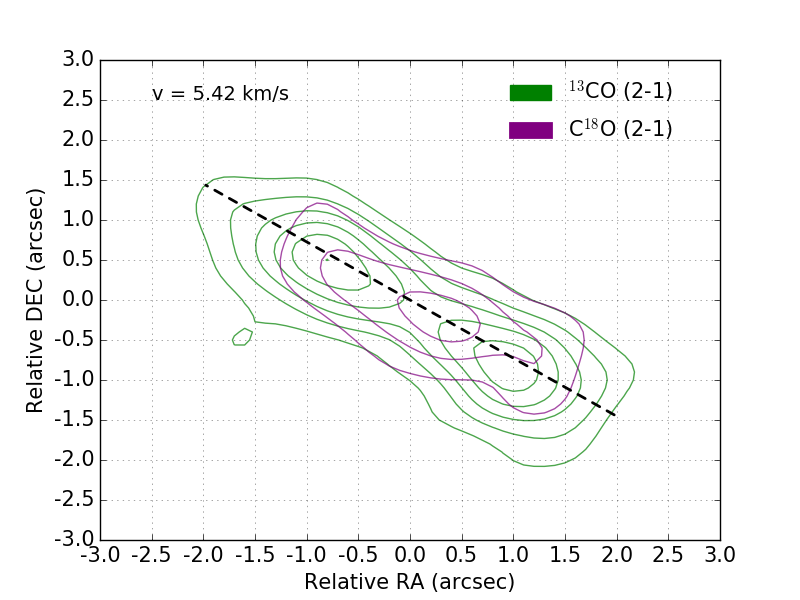}
\caption{Overlaid contour maps of the isotopologue channels at system velocity for determining disc position angle. Contours increase in multiples of 3 from 3$\sigma$, where for individual channels, $\sigma$= 23~mJy~kms$^{-1}$ /beam for $^{13}$CO and 11~mJy~kms$^{-1}$ /beam for C$^{18}$O. The dashed black line represents the adopted minor axis corresponding to a position angle of 144$^{\circ}$. }
\label{fig:pa}
\end{figure}

\section{Observations \& Data Reduction}
\label{sec:obs}
We present new ALMA observations of HD 100546 with band 6 receivers across 4 unique spectral windows (SPWs); project ID 2015.1.01600.S, PI: O. Pani\'c. The two continuum windows, SPW0 and SPW1, both contain 128 channels of width 15625 kHz to give a total bandwidth of 2~GHz, SPW0 has a central frequency of 232~GHz, with SPW1 at 234~GHz. 
SPW2 contains 1920 channels of width 61.035~kHz, which at central frequency 220398~MHz gives resolution of 0.083kms$^{-1}$ for the window containing the $^{13}$CO (2-1) emission line. 
The C$^{18}$O (2-1) line is in SPW3, this window has 3840 channels of 0.166~kms$^{-1}$ at central frequency 219560~MHz. Observations were made on the 28th March 2016 for a total time of 625s on source, employing 43 antennas with a baseline range of 15-460~metres giving a maximum angular scale of 10$\farcs$6. The flux calibrator was J1107-4449 and the phase calibrator was J1136-6827, the latter is found at the position 11:36:02.09787 -68:27:05821, which is (0$\farcs$22, 1$\farcs$73) degrees offset from the target.

Calibration and reduction was carried out using \textsc{casa} version 4.7.0. and imaging was performed using \textsc{casa} tasks and the \textsc{clean}  algorithm \citep{Rau2011}. Line free continuum channels were used to construct image models to be used for self-calibration, the solutions were then applied to all channels. After applying self-calibration, images were created with briggs weighting and robust=0.5, giving a synthesized beam of 1$\farcs$05~x~0$\farcs$77, P.A. = -20$^{\circ}$ for the continuum image. Source emission is contained within the inner third of the primary beam and so primary beam correction was not applied to images. Data cubes were created for the line emission by subtracting the continuum from SPWs containing the line emission using \textsc{casa} task \textit{uvcontsub}. Images of the lines have beam sizes of 1$\farcs$08 x 0$\farcs$82, P.A. = -19$^{\circ}$ for $^{13}$CO and 1$\farcs$10 x 0$\farcs$83, P.A. = -20$^{\circ}$ for C$^{18}$O. Integrated intensity, from which fluxes are measured, and intensity weighted velocity maps were created with \textsc{casa} task \textit{immoments} by summing all emission in channels containing significant (>3$\sigma$) signal; in the case of $^{13}$CO this corresponds to a range between -0.62 and 11.4 kms$^{-1}$ and for C$^{18}$O between -2.0 and 10.8 kms$^{-1}$.

We also utilise archival cycle 0 ALMA observations of HD 100546 in band 7 (Project ID: 2011.0.00863.S, PI: C. Walsh), as described by \citet{Walsh2014}. We re-reduce and image this data, applying self calibration as described above, to obtain continuum emission observed at a wavelength of 867$\mu$m. The cycle 0 data used J1147-6753 as a phase calibrator, offset from the primary target by (1$\farcs$2, 2$\farcs$3). The continuum image has a synthesized beam of dimensions 0$\farcs$95~x~0$\farcs$42, 38$^{\circ}$. The minimum baseline of 21 m results in a maximum angular scale of $5\farcs$3.

\section{Results}
\label{sec:results}
The integrated continuum flux in the 1.3~mm image of 492~mJy is measured with an image rms of 0.5~mJy/beam. 
The $^{13}$CO (2-1) integrated flux is measured at 12.872~Jy~kms$^{-1}$ and the first detection of C$^{18}$O in the disc gives an integrated flux of 2.948~Jy~kms$^{-1}$ for the (2-1) transition, with rms noise 0.033~Jy/beam~kms$^{-1}$ and 0.027~Jy/beam~kms$^{-1}$ respectively. The 2$\sigma$ contour reaches a maximum separation of 2$\farcs$5 in the 1.3~mm image, 3$\farcs$6 in $^{13}$CO and 3$\farcs$0 in C$^{18}$O. We discuss asymmetry and the radial extent of the emission maps thoroughly in Section \ref{sec:AD_asym}.

\begin{figure}
    \centering
    \includegraphics[width=\linewidth]{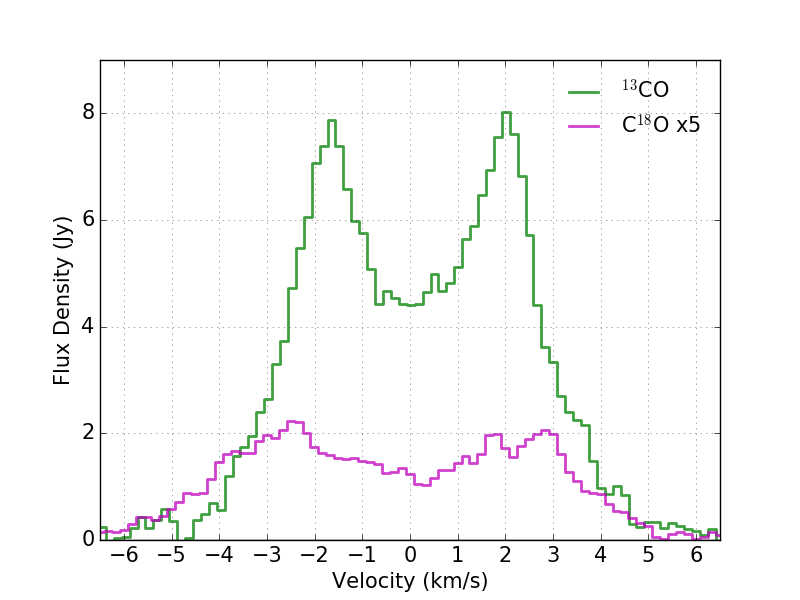}
    \caption{Spectrum from the spectral windows centred on the CO isotopologue lines. The C$^{18}$O line flux densities have been multiplied by a factor of 5 in order to make the profile easily visible in the plot. The x axis is measured relative to the system velocity of 5.42 kms$^{-1}$.}
    \label{fig:line_profs}
\end{figure}

To find a position angle for the source we use the data cubes from the CO isotopologue emission to identify the minor axis of the disc. We find the minor axis by locating the channel at the systemic velocity of the source, identified by the symmetry of the map as here it will show no Doppler shifting. Channel maps show the minor axis of $^{13}$CO in the channel at 5.51$\pm$0.08 kms$^{-1}$ and for C$^{18}$O at 5.33$\pm$0.17 kms$^{-1}$. We adopt an average of these two velocities as $v_{\rm sys}$ from this data set, giving 5.42 kms$^{-1}$, shown in Figure \ref{fig:pa} along with the minor axis channel of each isotopologue. This average is consistent with the two $v_{\rm sys}$ determined from the channel maps within the uncertainties (which we assume to be represented by the channel widths). 
Figure \ref{fig:line_profs} plots the observed spectrum of each of the two lines. Both lines display a double peaked spectrum, where each of the C$^{18}$O peaks is broader than the peaks of $^{13}$CO. The mid-point of the $^{13}$CO spectrum is consistent with the average $v_{\rm sys}$. The mid-point of the C$^{18}$O spectrum is harder to determine due to its shape. The local minimum at the centre of the double peak profile is offset from the adopted $v_{\rm sys}$ by 0.33 kms$^{-1}$. \citet{Walsh2014} constrain a system velocity of 5.7 kms$^{-1}$ from their ALMA observations of $^{12}$CO. 

The orientation of the disc minor axis is used to determine the position angle of the disc. We use our constrained position angle as an a priori estimate for a Gaussian fit of the data, in order to determine the inclination of the source. Any asymmetry in the emission map will influence the fitting of a Gaussian, by using this observationally determined a priori estimate for position angle, we can ensure our fitted inclination is more accurate.

For $^{13}$CO this method gives a position angle of the major axis, measured East of North, of $142\pm 0.8^{\circ}$ and for C$^{18}$O, $146\pm 2.3^{\circ}$. The average value of the two is $144\pm 2.4^\circ$ which we adopt henceforth. Uncertainties give the variation in position angle when measured from the adjacent spectral channels. Keeping this position angle fixed, we then fit an elliptical Gaussian to each of the CO isotopologue integrated emission maps to determine inclinations of $42.6^{\circ}$ and $43.7^{\circ}$ for $^{13}$CO and C$^{18}$O respectively. We adopt an average inclination of $43^{\circ}$ for all further analysis.

These values are consistent with most reported position angles from previous sub-mm observations of the disc with ALMA; \citet{Walsh2014} find a position angle of 146$^{\circ}\pm4^{\circ}$ and inclination 44$^{\circ}\pm3^{\circ}$, whilst \citet{Pineda2014} find 145.14$^{\circ}\pm0.04^{\circ}$ and 41.94$^{\circ}\pm0.03^{\circ}$. \citet{Panic2014} observed the disc with the MID-infrared Interferometric instrument (MIDI) on the VLT and find a position angle of 145$ \pm 5 ^{\circ}$, their inclination is considerably higher however because their observations probe the inner wall of the disc at 11~au. There is growing evidence for a warp in the disc \citep{Quillen2006,Panic2010Warm100546,Walsh2017} which would cause inclination to deviate radially. 

Asymmetry in the $^{12}$CO line profile is detected in both APEX and ALMA observations \citep{Panic2009CharacterisingLines,Walsh2017}. Images of the ALMA observations of $^{12}$CO achieve a synthesizing beam of 0$\farcs92$x0$\farcs$38 \citep{Pineda2014}. We do not detect this asymmetry in the less abundant $^{13}$CO or C$^{18}$O lines, finding instead double peaked line profiles symmetric to within uncertainties (Figure \ref{fig:line_profs}). 
$^{12}$CO emission traces higher scale heights in the disc, meaning that the asymmetry could be restricted to the upper vertical layers of the disc. There already exists evidence for asymmetry of the disc surface in the scattered light images that identify spiral arms and other structures \citep{Ardila2007,Avenhaus2014,Garufi2016,Follette2017}. The APEX observations have a larger beam and therefore the results may be affected by the extended envelope of the disc as detected by scattered light as far as $\sim 1000$~au \citep{Grady2001,Ardila2007}.

We adjusted the GAIA data release 2 (2018) position of HD~100546 for proper motion between the dates of the ALMA and GAIA observations, giving a position of 11h33m25.3209s -70d11m41.2432s, within an accuracy of (0.04, 0.05) mas \citep{Lindegren2018}. We find that the centre of emission for each of the tracers we observe is offset from this position. 

\begin{table}
\begin{tabular}{lll}
     & Coordinates of best  & Offset from Stellar  \\ 
     & fitting model       & Position ("), P.A. \\ \hline
Cont. (Im.)      & 11:33:25.305 -70.11.41.258 & 0.082,  79$^{\circ}$\\
$^{13}$CO (Im.)     &  11:33:25.310 -70.11.41.291 & 0.073, 49$^{\circ}$\\
C$^{18}$O (Im.)      &  11:33:25.305 -70.11.41.300 &0.098, 55$^{\circ}$\\
Cont. (uv) &   11:33:25.305 -70.11.41.257  & 0.082, 80$^{\circ}$\\
$^{13}$CO (uv) &  11:33:25.308 -70.11.41.276 & 0.072, 63$^{\circ}$\\
C$^{18}$O (uv)  &  11:33:25.300 -70.11.41.280 & 0.110, 71$^{\circ}$                      
\end{tabular}
\caption{Sexadecimal positions of central emission in the 1.3~mm continuum and CO isotopologues as determined by fitting a Gaussian to integrated emission maps in the image plane (`Im.' above), and a uniform disc in the uv plane (`uv'). Positions are rounded to the nearest milliarcsecond. The angular separation between these positions and the GAIA position of HD 100546, corrected for proper motion between GAIA and ALMA observations, is given in the final column.}
\label{tab:positions}
\end{table}

Table \ref{tab:positions} gives the coordinates of the centre of emission for the continuum and CO isotopologues, accompanied by its offset from the GAIA coordinates. A 2D Gaussian is fitted to the emission maps in Figure \ref{fig:ims}, the positions of these fits are taken to be the centre of emission. We also undertake a similar analysis, this time fitting a uniform, elliptical disc geometry in the uv plane using \textsc{CASA} task \textit{uvcontsub}. We repeat this for each tracer, giving the position of the best-fit disc model in Table \ref{tab:positions}. The benefit of uv-analysis is that it avoids being misled by spurious artefacts that can be introduced into images during the image reconstruction process. In each case the uv-fitted position is very similar to that determined in the image plane. The uniform-disc fits give a position angle of 146$^{\circ}$ and inclination of 42$^{\circ}$, values which are within a few degrees of those we constrain above, and those from previous ALMA observations \citep{Pineda2014,Walsh2014}.

The uncertainty in an astrometric position in an ALMA image depends on a number of factors, which we address here. The signal-to-noise based target position-fitting error is 1.3 mas. This is consistent with the ALMA technical handbook Section 10.6.6 .
The dominating factor in positional uncertainty is introduced when the phase corrections are transferred from the phase calibrator to the target. The phase calibrator is 1.7 degrees in separation from HD~100546, and from the phase drift between scans we estimate an total uncertainty on the astrometric position of $\sim$10 mas. This is relatively low as a result of a stable phase in the observations. Other factors such as antenna position errors can degrade astrometry; the ALMA technical handbook cautions that with reasonable phase stability, positional accuracy is $\sim$1/20 of resolution, which in this case would correspond to a maximum of 50 mas. We adopt this greater value of 50 mas as a conservative uncertainty on coordinate positions in the image. Each of the best-fit centres of emission in Table \ref{tab:positions} have a separation from the central star that exceeds the 50 mas uncertainty in ALMA image positions.
An offset of emission from the stellar position may indicate an over-density of disc material towards one side of the disc. We discuss asymmetry in the disc in more detail in Section \ref{sec:asym_images}. 

\section{Analysis and Discussion}
\subsection{Disc Mass}
\label{sec:AD_mass}

\subsubsection{Dust Mass}
\label{sec:dmass}

We measure the flux within a 3$\sigma$ contour of the 1.3~mm continuum image.  The opacity depends on the maximum grain size, for which we only have loose constraints. A minimum dust mass estimation can be made by assuming a dust grain size distribution  $dn/da \propto a^{-3.5}$ \citep{Mathis1977}. Assuming a maximum grain size $a_{\rm max}$~=~1~mm yields $\kappa_{1.3 \rm mm}$ = 1.15 $\rm cm^{2}g^{-1}$ \citep{Draine2006}.
Mid-plane temperature of the dust is difficult to constrain from observations, however we can get a firm handle on the mid-plane disc mass by assuming a sensible temperature range for this dense, cool environment, we  therefore adopt a range of 20K -- 40K. In this transition disc, gas and dust are likely to be well coupled, and our strong gas line detections suggest that the disc temperature must be over the CO freeze-out temperature of $\sim$20~K. Herbigs host more luminous, warmer discs than their T Tauri counterparts, and so the temperature is likely to be above this 20~K lower bound. We adopt an upper bound of 40~K. In the radiative transfer models of A type stars by \citet{Panic2017}, gas mass in the disc must be increased by over two orders of magnitude to increase mid-plane disc temperatures from 20 to 40~K at short separations from the star, and so we consider this to be a generous upper limit. 

\begin{figure}
\centering \includegraphics[width=0.9\linewidth]{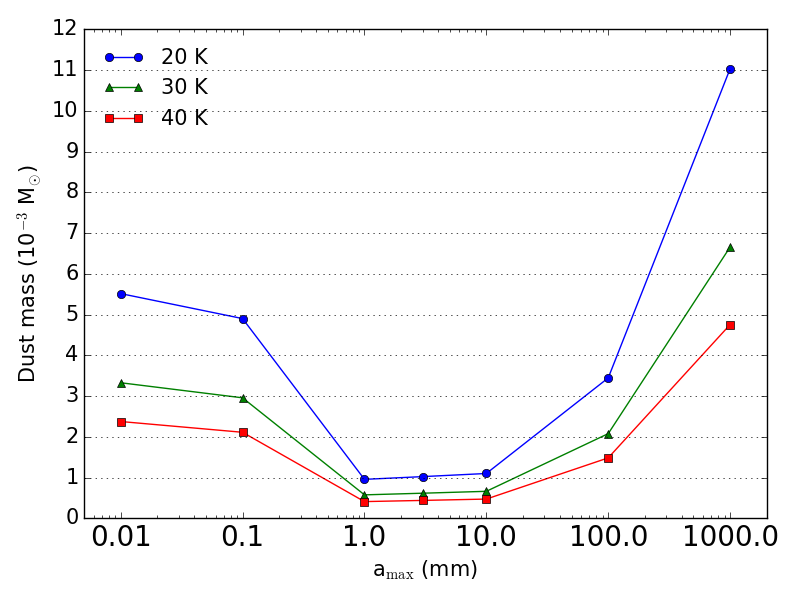}
\caption{Dust mass calculated over a range of temperatures and maximum grain sizes assuming optically thin emission measured at 1.3~mm.}
\label{fig:dustmass}
\end{figure}

Using Equation \ref{eqn:dust_mass} with the assumptions above, this gives a minimum total dust mass of 4.1$\times 10^{-4}~$M$_{\odot}$ for $T$=~40 K, and 9.6$\times 10^{-4}~$M$_{\odot}$ for $T$=~20 K. In Figure \ref{fig:dustmass} we show how this dust mass can vary with mid-plane temperature or if the assumed level of grain growth varies, thereby changing $a_{\rm max}$ and the emissivity of the dust. Between $a_{\rm max}$= 1~mm and 10~mm, the most likely values in a protoplanetary disc at this stage of evolution, the masses remain almost constant and are less sensitive to temperature. This indicates that for reasonable assumptions of $a_{\rm max}$ and mid-plane temperature, the dust mass is $\approx$10$^{-3}~$M$_{\odot}$.

The mass range calculated is comparable with the estimate of \citet{Wright2015} who employ the accretion disc models of \citet{dAlessio2005} to find M$_{\rm dust}$ of order 1.0$\times 10^{-3}~$M$_{\odot}$. Repeating the calculation above using instead the ALMA measured flux at 867$\mu$m from \citet{Walsh2014} with a temperature of 30K and $a_{\rm max}$~=~1~mm, a disc mass of 0.7$\times 10^{-3}~$M$_{\odot}$ is found, similar to the results between 1mm and 10mm in Figure \ref{fig:dustmass}.

This is more massive than most discs around low-mass stars, which fall in the range $10^{-6} - 10^{-4} \rm M_{\odot}$ \citep{Manara2018WhyPopulation}. This dust mass is also at the higher end of the range found for Herbig Ae stars $10^{-4} < M_{\rm dust} (M_\odot) < 10^{-3}$ (Pani\'c at al., in prep). Given the age of the star this is quite exceptional, suggesting a prime object to study longevity of dust in discs.

\subsubsection{Gas Mass}
\label{sec:gmass}

From the integrated line flux of C$^{18}$O above 3$\sigma$, assuming optically thin emission and local thermodynamic equilibrium, we calculate the total gas mass similarly to \citet{Hughes2008,Matra2015,White2016},
\begin{equation}
M_{\textrm{gas}} = \frac{4 \pi}{h \nu_{21}}  \frac{F_{21} m d^{2}}{A_{21} x_2},
\label{eqn:gas-mass}
\end{equation}
where F$_{21}$ is the integrated line flux, $\nu_{21}$ is the rest frequency of the transition, d the distance to the source, m is the mass of the CO molecule, $A_{21}$ is the appropriate Einstein coefficient, and $x$ is the fractional population of the upper level. We adopt ISM fractional abundances for CO and its isotopologues, specifically CO/H$_2\sim 10^{-4}$, and then adopt the relative fraction of C$^{18}$O in the ISM, CO/C$^{18}\rm O \sim$ 550 \citep{Wilson1994}. A gas mass is calculated in this way over a range of temperatures. The minimum gas mass was calculated with a temperature of 20~K, giving a minimum disc gas mass of 0.018~M$_{\odot}$. This estimated gas mass is around 1\% of the stellar mass, corresponding to around 18 Jupiter masses. 
For comparison, in an ALMA survey of Lupus, most low-mass discs have a gas disc mass of $\le 10^{-3} \rm M_{\odot}$. The gas mass found here is large even for Herbig stars, where the majority have discs within the range (1-10)$\times 10^{-3} M_{\odot}$ when calculated from mid-plane tracing CO isotopologues (Pani\'c at al., in prep). HD100546 therefore appears to still possess enough disc mass to form Jupiter mass planets, a very significant amount for a disc that has reached the transition disc stage of its evolution and shows detections of active planet formation \citep{Quanz2015CONFIRMATIONAU}.

The calculated gas mass assumes ISM abundances of C$^{18}$O relative to CO. Isotope-selective photodissociation can alter these ratios significantly in some discs, which can lead to underestimates of disc mass by up to an order of magnitude \citep{Miotello2014}. The CO line luminosities we measure in this data are unable to conclusively determine the extent of any isotope-selective photodissociation when compared to the the models of \citep{Miotello2016}, and so our calculation represents a lower limit. Carbon freeze-out beyond the snowline can also result in low measured gas masses; T Tauri discs appear to have significant levels of carbon freeze-out, for example in TW Hya \citep{Schwarz2016}. Herbig discs such as HD~100546 on the other hand remain warmer for a greater radial extent due to the more massive host star \citep{Panic2017}. This pushes the CO snowline to a larger radial separation allowing a larger amount of carbon in the disc to remain in the gas phase. In the disc of HD 100546 we do not expect freeze-out to play a significant role. Fitting physical-chemical models to observations, \citet{Kama2016} find that within the disc of TW Hya, C and O are strongly under-abundant, whereas in HD 100546, the depletion of gas-phase carbon appears quite low.

In light of the discussion above, our calculated gas estimate represents a firm lower limit, but a very large one in comparison to other Herbig discs. The gas mass estimated here is similar to that achieved through line modelling of another bright Herbig disc, HD 163296 \citep{Boneberg2016}, and shows there is still a large budget remaining in the disc for giant planet building in comparison to discs that are in a more advanced stage of their evolution towards a debris disc, e.g. HD 141569 \citep{White2016,Miley2018}.

\subsubsection{Gas To Dust Ratio}

Taking the dust mass calculated with $T$=20K and the minimum gas mass calculated at the same temperature, we calculate a g/d for HD 100546 of $\approx$ 19.
Low g/d (< 100) is observed in the majority of sources in recent surveys of discs in star forming regions \citep{Ansdell2016,Long2017}. Studies of individual Herbig discs suggest g/d is generally low in these larger, brighter discs too \citep{Meeus2010,Boneberg2016}. The g/d affects disc evolution because gas dynamics dominate the behaviour of the dust in gas-rich protoplanetary discs. The uncoupling of these two populations when gas densities drop sufficiently low is a signpost for the transition to a debris disc.

Low observed g/d can mean a great amount of gas has been lost, or it can be the result of carbon depletion in the disc through sequestration into large icy bodies or the conversion of CO into more complex molecules. This ratio is indicative only of regions within the CO snowline where freeze out is avoided. In the similar Herbig disc HD~163296 the snowline is calculated to be at around 90 au \citep{Boneberg2016}, in cooler discs the snowline will be at a lesser separation, e.g. TW Hya, 40 au \citep{Schwarz2016}.

\subsection{Disc Asymmetry}
\label{sec:AD_asym}

\subsubsection{Evidence in ALMA Images}
\label{sec:asym_images}


\begin{figure}
\centering \includegraphics[width=0.99\linewidth]{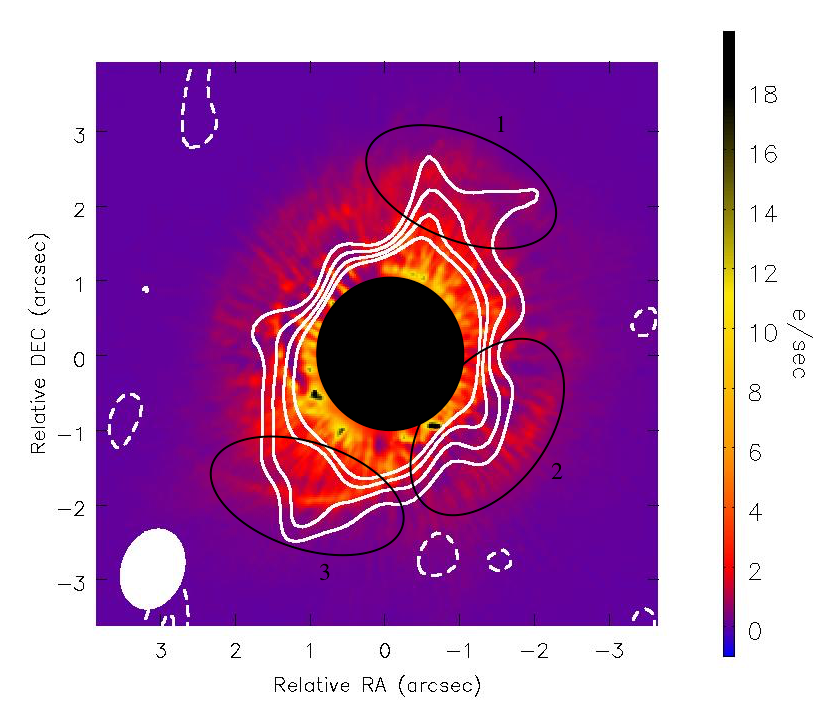}
\caption{The colour image shows the PSF deconvolved HST ACS image of HD~100546 in the F435W band. Overlaid in white contours is the integrated C$^{18}$O (2-1) emission observed by ALMA, with contours at (-2,2,3,4,5)$\times \sigma_{18}$, dashed lines are negative contours. Structures in the scattered light image initially identified by Ardila et al. are indicated with black ellipses. The black circle of radius 1$\farcs$0 represents the HST coronograph. }
\label{fig:hst}
\end{figure}

Asymmetry in the disc has previously been reported with evidence in asymmetric $^{12}$CO line profiles \citep{Panic2009CharacterisingLines}, horse-shoe structures in cm-wavelength observations \citep{Wright2015} and the identification of multiple scattered light features \citep[e.g.][]{Ardila2007,Garufi2016,Follette2017}. 
Asymmetric features can form in the disc as a result of mechanisms such as planet-disc interactions \citep{Baruteau2014}, interactions between binary companions \citep{Price2018} or particle traps within the disc \citep{vanderMarel2013ADisk,VanDerMarel2016}. Such features are therefore particularly interesting in a disc such as that of HD~100546 which is known to host an embedded protoplanet, and exhibits a range of observed structures across different spatial scales.


In the data presented here we see a striking match with the extended emission in the C$^{18}$O map and the structures in HST scattered light observations of the disc. The low-level bulges in C$^{18}$O emission (Figure \ref{fig:hst}) persist in images created using different weighting parameters in the image reconstruction process, and are present both before and after applying self-calibration.
Using the Advanced Camera for Surveys (ACS) on board the Hubble Space Telescope (HST) \citet{Ardila2007} identify 3 structures in the outer disc which are attributed to the known spiral arms in the disc. The authors name these structures 1, 2 and 3 and we adopt this convention here as well. Figure \ref{fig:hst} displays the HST image in the F435W band presented by \citep{Ardila2007}. 
Each of the low-level bulges in the extended C$^{18}$O (2-1) emission is found in the same region of the disc as one of the 3 scattered light structures. This provides tentative evidence of a mid-plane gas equivalent to the scattered light features.
The C$^{18}$O emission is optically thin, and can therefore be interpreted as an increase in local mid-plane density. An increased local gas density will increase the scale height of the disc, the subsequent elevation of small particles towards the scattering surface will result in a stronger scattered light flux, as we see in the HST data. The C$^{18}$O emission bulges towards the outer disc may therefore be signalling the mid-plane density features that correspond with the disc's spiral arms.
In this section we make a general analysis of any observational evidence for asymmetry in the disc, within the limits of the angular resolution of our ALMA observations.

\begin{figure}
\centering \includegraphics[width=0.99\linewidth]{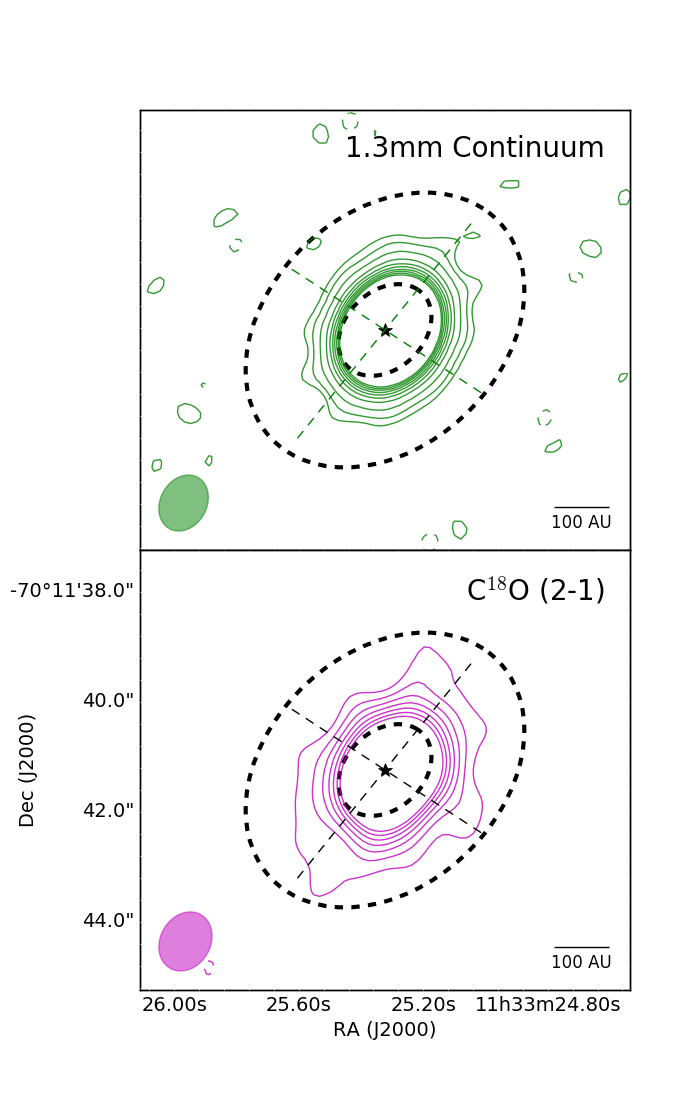}
\caption{Emission maps of 1.3mm continuum and C$^{18}$O (2-1) with overlaid dashed ellipses with a position angle of 144$^{\circ}$ and inclination of 43$^{\circ}$. These correspond to rings with a deprojected radius of 100~au and 300~au. The star marks the GAIA position of HD 100546 on which the ellipses are centered. Contours in the top panel range from (5-75)$\times\sigma$ in steps of 10$\sigma$ and in the bottom panel from (3-15)$\times\sigma$ in steps of 2$\sigma$. Dashed contours are negative. High level contours internal to the ellipse are omitted for clarity.}
\label{fig:asym_ellipses}
\end{figure}

In Section \ref{sec:results} we accurately measure the inclination and position angle, both of which are very close to values derived from previous sub-millimetre ALMA observations of the disc \citep{Pineda2014,Walsh2014}. We take this geometry to construct an ellipse with a semi-major axis equivalent to 100~au in the disc. We can make an initial visual assessment of asymmetry in the disc by plotting this ellipse over the emission maps in Figure \ref{fig:ims} centred at the stellar position. The mid-plane tracing 1.3~mm continuum and C$^{18}$O (2-1) emission in particular show evidnece on an offset disc in Figure \ref{fig:asym_ellipses}, on scales larger than that the offsets listed in Table \ref{tab:positions}. In both panels we omit high level contours internal to the inner ellipse for clarity. Towards the northwest in the panels of Figure \ref{fig:asym_ellipses}, the 75$\sigma$ contour (in the 1.3~mm image) and the 15$\sigma$ contour (C$^{18}$O) overlap with the 100~au ellipse, whereas to the southeast there is a significant offset. C$^{18}$O in particular shows azimuthal asymmetry in the outer disc, with low level emission extending much outward as previously highlighted in Figure \ref{fig:hst}.

\begin{figure}
\centering\includegraphics[width=\linewidth]{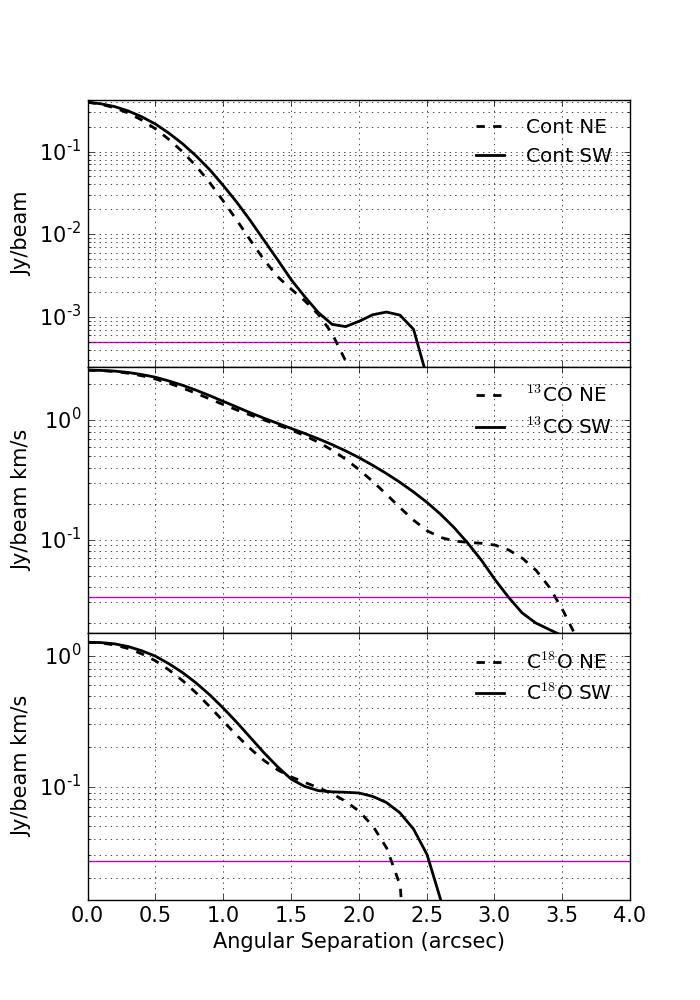}
\caption{Slice are extracted from the emission maps of each tracer along the minor axis. Solid lines show the slice towards the southwest, dashed lines show the slice to the northwest. The horizontal purple line indicates the rms measured in each image. }
\label{fig:slices}
\end{figure}

The largest asymmetry in the disc is along the minor axis (Figure \ref{fig:asym_ellipses}) with the brighter region found towards the position of Structure 2 in Figure \ref{fig:hst}. We can also see this in Figure \ref{fig:slices}, which shows slices from the emission maps taken along the minor axis in both directions, centered at the stellar position.

C$^{18}$O (bottom panel) is brighter in the southwest for the majority of the disc, and emission extends further by $\sim$0$\farcs$3 in comparison to that in the northeast. 
In the continuum (top panel) the southwest side is brighter than the northeast throughout the disc. The greatest difference between the two sides is at an angular separation of 0$\farcs$7, where the southwest is brighter by 50$\sigma$. A low-level bump in the continuum emission just beyond 2$\farcs$0 occurs at the same radial location as the extended C$^{18}$O emission and scattered light emission in Figure \ref{fig:hst}. 
$^{13}$CO emission (middle panel) is also brighter in the southwest for as far as $\sim2\farcs$8, beyond which the northeast of the disc is brighter.

\subsubsection{Comparison to previous asymmetries in HD 100456}

Figure \ref{fig:hst} displays how the low-level, extended emission towards the edge of the gas disc is spatially coincident with structures identified by \citet{Ardila2007}, potentially linking observed scattered light features with the mid-plane gas above which the small, scattering grains are suspended. In other scattered light observations, \citet{Avenhaus2014} see a dark lane that extends from  $\sim 0\farcs2 - 0\farcs6$ along the minor axis. The authors attribute this to effects due to the polarised scattering function, rather than any physical feature in the disc. However \citet{Wright2015} also observe an arc of millimetre emission in the same direction, at the same separation as the scattered light dark lane, but not quite as extended. 

Mid-IR interferometry is able to resolve the inner disc, where asymmetry is found close to the inner gap of the disc at 9 au \citep{Panic2014}. Features in the the inner disc can have a strong effect on the outer disc.
For example there has been recent evidence of warps from ALMA observations, e.g. this disc \citep{Walsh2017} and T Tauri star AA Tau \citep{Loomis2017}. A warp of the inner disc can lead not only to shadowing of the outer disc but potentially even a break of the disc into two distinct planes. Substructure can then develop as a result of this geometry; a misaligned disc can result in shadows that create asymmetric illumination of the disc, whilst a `broken' disc can contain dust streamers that cross gaps in the disc \citep{Dutrey2014PossibleA,Loomis2017}. Recent very high angular resolution (0$\farcs05$) ALMA observations of the disc resolve the very inner regions of the disc, revealing an asymmetric circumstellar ring in the continuum, brightest towards the North East  \citep{Pineda2018High-ResolutionLimits}. The authors attribute this asymmetry to another scattered light spiral feature, this time identified by SPHERE \citep{Garufi2016}, on angular scales much less than that of the synthesized beam achieved in the observations presented here.

Many of these features can arise through interactions with a central companion as shown by 3D hydrodyanamic simulations of the transition disc HD~142527 \citep{Price2018}, this is particularly interesting in regard to any asymmetries found in the HD~100546 disc when bearing in mind the evidence found in recent years for a further companion at smaller separation.

\subsection{Asymmetry Analysis}
\label{sec:asym_anal}

Asymmetry in the surface brightness of the gas emission can occur from increased temperature, increased density or a combination of the two.

In 3D hydrodynamic simulations, spiral arms launched by planets observed at 1.3~mm have been shown to induce azimuthal variation in gas surface density of approximately 50-60\% \citep{Juhasz2018}, and should also be observable in optically thin gas observations. In this scenario, a large spiral arm could increase gas density locally, thereby elevating the scale height of micron-sized grains to produce the scattered light features. Alternatively a local density increase of 30\% over a length scale$\approx$ disc scale height is enough to provide the conditions for dust trapping of larger particles in the outer disc \citep{Pinilla2012}. Another explanation for local increase in density is a gravitationally unstable disc, where clumping occurs due to gravitational forces if the free fall time scale is short compared to sound speed crossing time scales and shearing time scales \citep{Toomre1964}. This is generally found in discs at an earlier stage in their evolution whilst discs are still very massive $(\rm M_{disc}/M_{\star} > 10\%$), which is not the case here.

Brightness asymmetry can also result from varying temperature in the disc. Azimuthal variation in disc temperature may result from an inner warp that obscures sections of the disc or azimuthal variation of density.

\subsubsection{Density Scenario}
\label{sec:density_scenario}

In this scenario we assume that the difference in observed brightness is due solely to changes in gas density. We split each half of the disc, bisected along the major axis, into three 60$^\circ$ degree segments and sum the flux in the segment. By calculating the brightness ratio between two segments on opposite sides of the disc, we can calculate the increase in surface density required to produce the difference in flux. 
The intensity of integrated line emission is given by
\begin{equation}
    I =\frac{h \nu A_{J,J-1} N(J)}{4 \pi},
\label{eqn:I}
\end{equation}
where A$_{J,J-1}$ is the Einstein coefficient for spontaneous emission from level J to level J-1, and N(J) is the column density in a specific level. The intensity is therefore proportional to the density of molecules in the N(J) level.

The largest brightness difference is between the two segments along the minor axis; the southwest is brighter than the northeast by a factor of 1.10$\pm$0.15 . The asymmetry identified in Section \ref{sec:asym_images} is most obvious in the outer disc. We also compare the brightness ratio of the outer disc, we do this in the same fashion, except only summing the emission beyond a deprojected radius of 100~au. In this case, the brightness ratio is 1.27$\pm$0.18 . We are limited to comparisons that average over regions of the disc due to the angular resolution of our observations, but local density increases on smaller scales can alert us to certain physical mechanisms in the disc, some of which we discuss below. 

Given the 1$\farcs$0 resolution of the data, it is feasible that there are much larger azimuthal density variations for a given radial separation that are diluted by the beam. This might arise due to a local increase in density due to, for example, a spiral arm. \citet{Ardila2007} associate the three structures in Figure \ref{fig:hst} with spiral arms, and it is towards each of these structures that the outer disc of C$^{18}$O emission seems least symmetric.

From this data we cannot rule out the possibility of an unresolved vortex capable of trapping particles at large separation in the HD~100546 disc, as our density variation is similar to that predicted by \citet{Pinilla2012}. High resolution data across multiple wavelengths would help to identify any such feature \citep[e.g.][]{Cazzoletti2018}.  

Calculation of the Toomre parameter, Q, based on the radial profile of C$^{18}$O rules out gravitational instability in the disc. Assuming C$^{18}$O traces total disc mass, Q(R) for this disc reaches a minimum $\sim 24$ at $\sim$80~au confirming that the disc is stable against gravitational collapse. However a clump in the outer disc will experience relatively long orbital timescales, meaning it will take longer to break up. Fragments of the disc that were previously unstable might linger at the outer edges of the disc as they dissipate.

\subsubsection{Temperature Scenario}

To test the possibility of temperature variation as the cause of asymmetry in the outer C$^{18}$O disc, we measure the average flux in different segments of the disc as in the previous section, and then calculate the temperature variation that would be required to reproduce such a flux difference. Through a comparison of the average fluxes in opposite segments, we can calculate the increase in temperature required if this was to be due to temperature alone (i.e. no density increase).

We calculate the column density of C$^{18}$O gas in the J=2 level using 
\begin{equation}
    N_{J}= N(total) \frac{2J+1}{Z} \frac{1}{exp\big(\frac{h B_e J(J+1)}{k T}\big)},
\label{eqn:Ntot}
\end{equation}

where T is the midplane temperature and B$_{\rm e}$ is the rotation constant. The partition function, Z, is calculated using
\begin{equation}
    Z = \sum_{J=0}^{\infty}(2J+1) ~exp\Big(\frac{-h B_e J (J+1)}{k T}\Big),
\label{eqn:Z}
\end{equation}

Substituting for Z and N(J), Equation \ref{eqn:I} becomes
\begin{equation}
I = \frac{h \nu A_{J,J-1}}{4 \pi} \frac{N_{\rm total}  ~(2J+1)}{exp\Big(\frac{h B_e J(J+1)}{kT}\Big)} \frac{1}{ \sum_{J=0}^{\infty}(2J+1) ~exp\Big(\frac{-h B_e J (J+1)}{k T}\Big) }.
\end{equation}

\begin{figure}
    \centering
    \includegraphics[width=\linewidth]{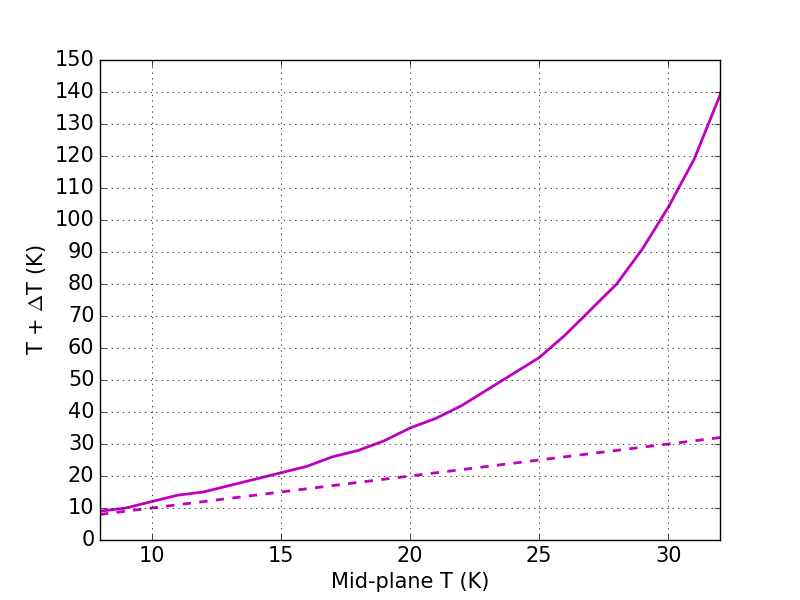}
    \caption{Calculated temperature, T+T$\Delta$T, required to produce the observed variation in flux in segments of the disc from the C$^{18}$O moment 0 map, over a range of assumed mid-plane temperatures. The dashed line indicates $\Delta T$=0.}
    \label{fig:gas_temps}
\end{figure}

Assuming a range of mid-plane temperatures, we then calculate $\Delta$T required to produce the observed brightness ratio of 1.27 from Section \ref{sec:density_scenario}. We plot the results against the range of assumed mid-plane temperatures in Figure \ref{fig:gas_temps}.

Figure \ref{fig:gas_temps} shows that even for a mid-plane temperature as cool as 20K (below which CO is no longer in the gas phase), a 75\% increase in average temperature is required. From 24K and above, the average temperature of the outer disc mid-plane must increase by greater than 200\%. There are no known physical processes expected within the disc mid-plane which could produce such a large change in temperature, especially when averaged over such a large proportion of the disc. If we were to assume that inner parts of the disc dominate the flux in each segment, temperatures may be larger than in the cool outer disc. However, the radiative transfer models of \citet{Panic2017} show that in reality an increase in disc temperature from 20 to 32K (comparable to the 20-35K increase in Figure \ref{fig:gas_temps}), requires an increase in gas density of order 10, far greater than what we detect here. We can therefore rule out temperature as a sole cause of any flux asymmetry. In order to explain the difference in brightness between the two regions of the disc we must invoke some sort of density increase as well.

\subsection{Radial Flux Profiles }

\label{sec:rad_profiles}
\begin{figure}
\centering 
\includegraphics[width=1\linewidth]{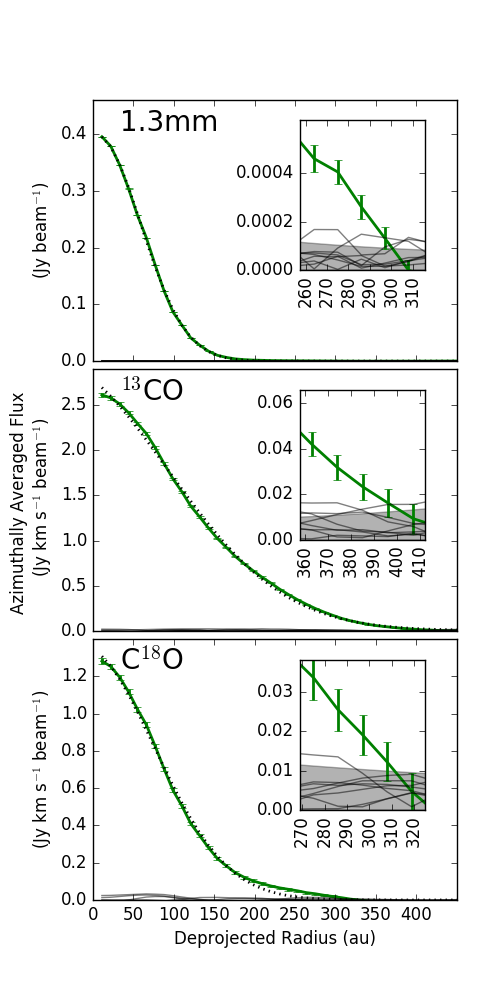}
\caption{Azimuthally averaged flux plotted against radius in green for both the continuum and line emission. Grey lines show the averages obtained for regions off-source (no emission). The shaded area represents the 2$\sigma$ level computed from these off-source averages. Dotted black lines are least-squares fitted Gaussians to the flux profiles. }
\label{fig:azi_avgs}
\end{figure}

Ignoring azimuthal variations in disc emission, we can consider how the flux changes in the radial direction by de-projecting and then azimuthally averaging, around the mm continuum centre, the emission from images in Figure \ref{fig:ims}. The resultant average profiles are shown in Figure \ref{fig:azi_avgs}.

Taking advantage of the high signal to noise in the data we adopt the following method to obtain an accurate uncertainty estimate for the radial flux profiles.

Each image is averaged off-source in `empty' sky locations at 45$^{\circ}$ position angle intervals at a separation that does not include any significant emission from the central source, but is also well within the primary beam. The resultant radial profile gauges the level of fluctuations due to background noise in that part of the image (shown as grey lines in Figure \ref{fig:azi_avgs}, more visible in the insets). The amplitudes of these fluctuations as a function of radius are fitted with polynomials, the polynomials are then averaged over the different off-source locations. The resulting final polynomial gives the average level of off-source fluctuations in the image as a result of the azimuthal averaging process. We use this for error bars on the average flux curve. Our method also allows for a determination of average outermost detection radius that does not rely on a pre-defined function or model, we take this to be the point at which the average flux becomes less than 2 times the average noise fluctuations at that radius; in Figure \ref{fig:azi_avgs} this corresponds to the error bars reaching the grey shaded area. 

Figure \ref{fig:azi_avgs} plots the azimuthally averaged continuum flux and integrated line emission against deprojected radius (using inclination=43$^{\circ}$ and PA=144$^{\circ}$ as determined in Section \ref{sec:results}). The 1.3~mm continuum for as far as an average, deprojected radial separation of 286$\pm$11~au. The $^{13}$CO profile shows us that the gas in the disc is significantly more extended; detected as far out as 385$\pm$11~au, a comparable distance to the 390$\pm$20~au extent of $^{12}$CO in the disc reported by \citet{Walsh2014}.

The less abundant molecule C$^{18}$O is detected out to a shorter separation of 297$\pm$11~au. Found at lower scale-heights, this molecule traces the cool, dense mid-plane \cite{VanZadelhoff2001,Dartois2003} in which giant planets are expected to form. 

A large amount of mass spread over large radial separations is important for planet formation in the disc. Giant planets grow more massive in discs of higher initial disc (gas) mass \citep{Mordasini2012AstrophysicsLifetime}. Furthermore, population synthesis models show that significant amounts of giant planets migrate into the central star after formation \citep[See review][]{Mordasini2018}. In order to survive, some cores accrete late in the disc lifetime and grow as they migrate inward \citep{Baruteau2014}. Our results suggest the disc of HD~100546 is well placed to support such late survivors, because a large amount of planet building material ($\sim$18 M$_J$) still exists over a large region of the disc (297~au outermost detection of C$^{18}$O). This is particularly relevant in the case of this disc, where a confirmed proto-planet lies towards the outer disc at a radial separation of $\sim$50~au \citep{Quanz2015CONFIRMATIONAU}. 

\begin{figure}
\centering\includegraphics[width=1\linewidth]{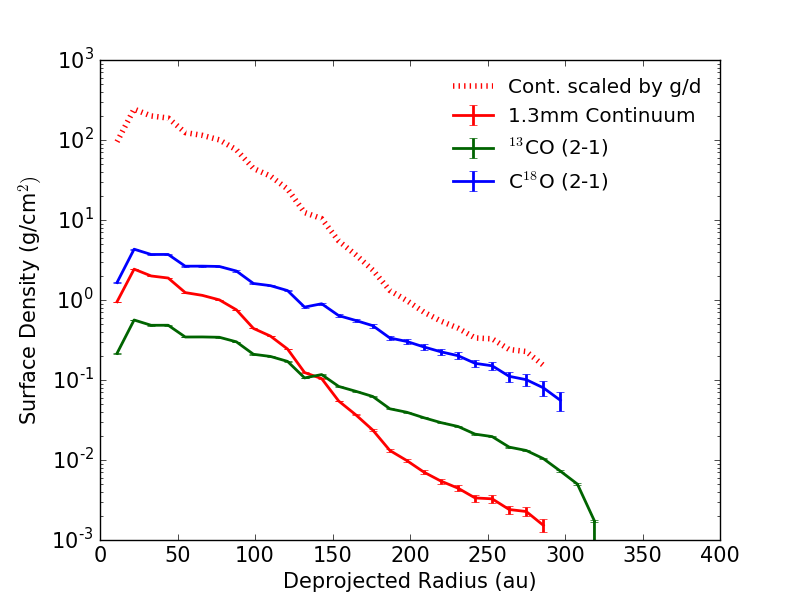}
\caption{Surface density profiles calculated form the radial profiles in Figure \ref{fig:azi_avgs}. The dotted line represents the surface density of the 1.3~mm continuum image scaled up by a factor g/d=100.}
\label{fig:surf_dens}
\end{figure}

The difference in the outer radius at which the discs are detected may be a true physical difference, or could be as a result of the different sensitivity between the continuum and the line emission. 
We explore this by calculating surface density profiles for each tracer, which are plotted in Figure \ref{fig:surf_dens}. 

In Figure \ref{fig:surf_dens} we plot the surface density profile of each tracer in order to assess how it changes with radial separation. The $^{13}$CO curve represents a lower bound because the emission is optically thick for most of the disc. The continuum surface density profile is scaled by an assumption that g/d in the disc is the canonical value of 100, giving an upper bound with which we can compare to the gas. The profiles in Figure \ref{fig:surf_dens} tend to show a steady decline towards the outer disc, rather than a sharp fall that might signify a definitive outer edge of the disc.

The surface density of mm particles, as calculated from the 1.3~mm emission, is fairly large to begin with but begins to drop sharply after $\sim$100~au. The C$^{18}$O surface density declines over the extent of the disc on the log scale in Figure \ref{fig:surf_dens}, converging with the g/d scaled continuum curve. The mass calculations in Section \ref{sec:results} result in a disc g/d lower than 100. If values of <100 are used to scale the 1.3~mm surface density curve, in the outer disc C$^{18}$O would soon have a greater surface density than the scaled dust, suggesting we have enough sensitivity in our gas line observations to confirm that the gas is more extended than the dust. Determining the outer radius of a disc in either gas or dust is complex, but \citet{Facchini2017} use thermo-chemical models to find that the observed difference between dust and gas disc radii is largely down to optical depth effects.  The effect of radial drift is shown by a sharp outer edge to the dust disc. Such a feature is absent from the dust surface density profile in Figure \ref{fig:surf_dens} due to the $\sim$100~au resolution in these observations.

\begin{figure}
\centering\includegraphics[width=0.9\linewidth]{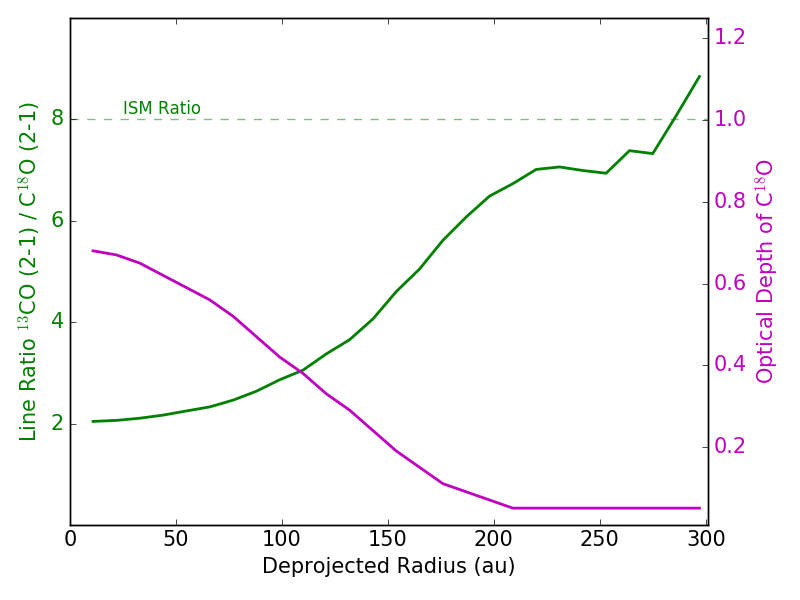}
\caption{The ratio $^{13}$CO/C$^{18}$O is plotted against de-projected radius and coloured in green, the calculated optical depth of C$^{18}$O as a function of radius is plotted in purple.}
\label{fig:lines_tau_ratio}
\end{figure}

We see evidence of optical depth effects in this data set in the CO isotopologues; throughout the disc $^{13}$CO has a lower surface density than the much less abundant C$^{18}$O. In order to estimate the optical depth of these species we use Equation \ref{eqn:gas_opt_depth},  
\begin{equation}
\frac{S (^{13}\rm CO)}{S (\rm C^{18}O)} = \frac{\nu_{^{13}\rm CO}}{\nu_{\rm C^{18}O}} ~ \frac{T_{ex,^{13}\rm CO}}{T_{ex,\rm C^{18}O}} ~ \frac{1-e^{\tau_{^{13}\rm CO}}}{1-e^{\tau _{\rm C^{18}O}}} .
\label{eqn:gas_opt_depth}
\end{equation}
This is a method adopted from \citet{Schwarz2016} where we have used the ratio of fluxes instead of brightness temperatures. We assume $^{13}$CO to be optically thick and the isotopic ratio $^{13}$CO/C$^{18}$O=8 based on ISM values \citep{Wilson1999}. The ratio of excitation temperatures of the two isotopologues is assumed to be $\approx$ 1. Figure \ref{fig:lines_tau_ratio} shows the resulting $\tau$(R) as well as the flux ratio from the two isotopologues. 

At the outer edge of C$^{18}$O detection, the ratio of isotopologue emission rises up to $\sim$7, with a steep increase towards the ISM ratio of 8 from $\sim$250~au as the edge of the C$^{18}$O disc is approached. C$^{18}$O is found to be optically thin for the full radial distance over which we detect it, with $\tau \le$ 0.68 as shown in Figure \ref{fig:lines_tau_ratio}. 
This optical depth result can help to explain the isotopic ratio curve, which rises as the optical depth decreases. Assuming the intrinsic isotopic abundance ratio within the disc remains constant, the $^{13}$CO optical depth will decrease as the C$^{18}$O does with separation from the star. As $^{13}$CO becomes increasingly optically thin, the observed abundance ratio will tend towards its true value.

\subsubsection{Spectral Index}
\label{sec:spec_ind}

\begin{figure}
\centering\includegraphics[width=\linewidth]{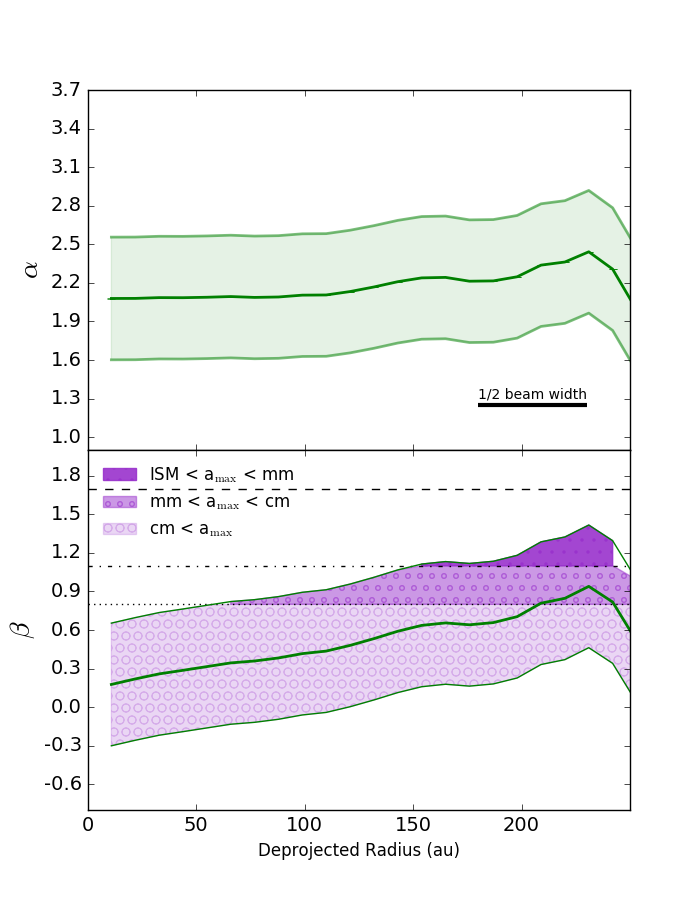}
\caption{Upper panel : Radially dependent spectral index, $\alpha$, calculated using ALMA continuum observations at 867 $\mu$m and 1.3 mm for the disc of HD 100546. The upper and lower green lines give the maximum and minimum curve considering flux calibration uncertainties. The bar indicates a length of 50au, approximately half the beam width. Lower panel: Radially dependent opacity index, $\beta$, over-plotted with horizontal lines that show the opacity calculated by \citet{Draine2006} for a dust population with $a_{\rm{max}}$ of cm- (dotted), mm- (dash-dot) and ISM grains (dashed), assuming a grain size distribution of $dn/da \propto a^{-3.5}$. Where the lower-bound curve crosses these lines, the area under the curve is coloured according to the key. Error bars are propagated from noise in the image (insets of Figure \ref{fig:azi_avgs}), and are barely visible. The x-axis is shown up to 230~au, the detected size of the 867$\mu$m disc \citep{Walsh2014}.} 
\label{fig:spec_index}
\end{figure}

We now utilise the 867~$\mu$m continuum data from the ALMA band 7 data on this source \citep{Walsh2014}. The continuum from this data is imaged using the same beam as the band 6 1.3mm continuum.
Using azimuthal averaged profiles of each continuum image we can directly calculate an average, radially dependent spectral index for the disc, plotted as the central green line in Figure \ref{fig:spec_index}. In the upper panel of Figure \ref{fig:spec_index}, $\alpha_{\rm mm}$(R) begins in the inner disc with a value of 2.1 and steadily rises with radius by $\Delta \alpha_{\rm mm} \approx$ 0.3 .

\begin{figure}
\centering\includegraphics[width=\linewidth]{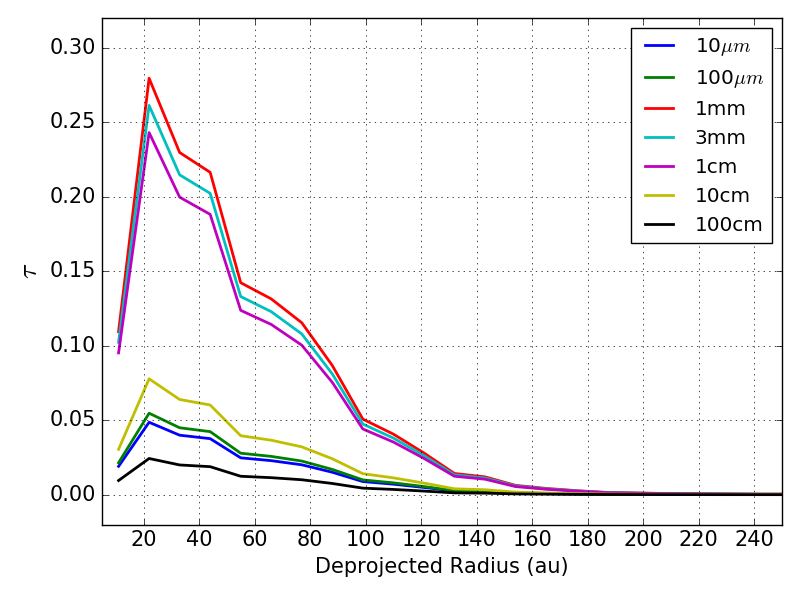}
\caption{Optical depth of the 1.3~mm emission as a function of de-projected radius calculated for a range of maximum grain sizes ($a_{\rm max}$) in the mrn grain size distribution.}
\label{fig:opt_depth}
\end{figure}

$\alpha_{\rm mm}$(R) remains within the range typically seen in protoplanetary discs \citep{Ricci2012,Pinilla2014}, covering a range of values that are low in comparison to that of the ISM. Low spectral index at millimetre wavelengths can be down to one of two reasons; grain growth has lead to larger dust particles ($a_{\rm max} \ge 3\lambda$) which emit less efficiently and reduce the calculated $\alpha$, or optically thick emission. As we do not have robust information on the sizes of grains present in the disc, we calculate the optical depth of 1.3 mm continuum emission as a function of radius using a range of emissivities at 1.3 mm corresponding to different particle sizes from \citet{Draine2006} . 
The results are shown in Figure \ref{fig:opt_depth} and suggest that over all emissivities tested, the 1.3~mm emission is always optically thin for the entirety of the disc. Through comparing spectral indices calculated from previous mm observations, \citet{Walsh2014} find that at sub-mm wavelengths the continuum emission may be approaching the optically thick regime however.

In order to constrain what the size of the emitting grains are we calculate the opacity index $\beta_{\rm mm}$. In an optically thin disc under the Rayleigh-Jeans approximation, this can be related to the spectral index by $ \alpha_{\rm mm} = \beta_{\rm mm} + 2$. At $\lambda \approx$ 1 mm we trace the cooler dust, where the Rayleigh-Jeans approximation does not hold, and so we take an approach similar to that of \citet{Guidi2016} and calculate $\beta_{\rm mm}$ with 
\begin{equation}
\beta_{\rm mm}(R) = \frac{ \rm d log \textit{F}_\nu \textit{(R)} - d log \textit{B}_\nu}{\rm d log \nu}.
\label{eqn:beta_calc}
\end{equation}

\begin{figure}
\centering\includegraphics[width=\linewidth]{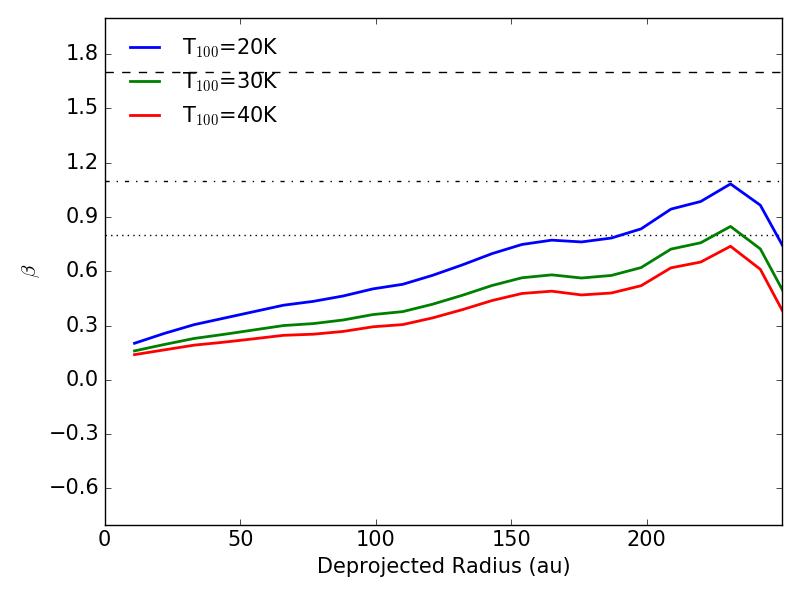}
\caption{$\beta_{\rm mm}$ calculated using temperature profiles calculated with a power law with varying T$_{100}$. Horizontal lines are the same as in the lower panel of Figure \ref{fig:spec_index}.}
\label{fig:beta_temps}
\end{figure}

We adopt a temperature profile that obeys a power law, $ T(R) = T_{100}~( R/R_{100} )^{0.5} $, and take $T_{100}$=25K. Figure \ref{fig:beta_temps} demonstrates how temperature affects the calculated $\beta_{\rm mm}$(R). Although temperatures in the mid-plane and the outer disc are cool and relatively stable, Figure \ref{fig:beta_temps} shows that an inaccurate temperature prescription can alter the $\beta$ significantly; a difference of 20K in $T_{100}$ changes $\beta$ at $\sim$200~au by 0.32. For comparison, this difference is larger than the difference between the characteristic $\beta$ for $a_{\rm max}$=1mm and $a_{\rm max}$=1cm. Under-estimating disc temperature increases the inferred size of grains in the cooler outer disc, meaning the radial size sorting of grain sizes is of a lesser degree and that grain sizes thoughout the disc are of large ($\sim$mm) sizes.

In the lower panel of Figure \ref{fig:spec_index} we show $\beta_{\rm mm}$ as a function of radius compared with opacity indexes calculated by \citet{Draine2006}. Coloured areas on the plot show radii at which $\beta_{\rm mm}$ indicates a particular $a_{\rm max}$ has been achieved through grain growth, determined by when the lower bound of the calculation crosses the indicative horizontal lines. The curve takes a similar form to that in the top panel, except for a larger variation of $\Delta \beta_{\rm mm}(R) \approx$ 0.8 over the extent of the disc. 

The main result from Figure \ref{fig:spec_index} is that $\beta_{\rm mm}(R)$ calculated between 867$\mu$m and 1.3~mm indicates grain growth to cm-sizes throughout the disc. In fact, the characteristic $\beta$ for $a_{\rm max} = 1,10,10^{2}$ cm, at millimetre wavelengths are all $\approx 0.8$ \citep[see][Figure 2]{Draine2006}, so dust could be growing even larger than cm sizes without us being able to detect it. Within the uncertainties due to flux calibration, this result is consistent with grain growth to cm-size for the full extent of the disc detected at both continuum wavelengths, where we calculate $\alpha$ over the shorter $\sim 230$~au band 7 continuum radius \citep{Walsh2014}. $\beta_{\rm mm}(R)$ for the disc inward of $\sim$200~au is consistent with pebble (>cm) sized dust. The upper curve remains below the top dashed line for the full extent of the disc, confirming that grains have evolved from an ISM-like $\beta$ throughout the disc. Overall, mid-plane dust has grown substantially from ISM levels within the disc.

Radially increasing $\beta_{\rm mm}$ in the outward direction has been predicted by dust evolution models that include realistic coagulation and fragmentation \citep{Brauer2008,Birnstiel2010b}. Radially increasing $\beta$ also agrees with past studies that fit multi-wavelength observations with parametric models of discs in the uv-plane, where only those with a radially increasing $\beta$ could reproduce observed emission \citep{Banzatti2011AstronomyTauri,Perez2012CONSTRAINTSDISK,Trotta2013,Tazzari2016}. 

\citet{Guidi2016} use high resolution data from ALMA and the VLA to observationally confirm a radially increasing dust opacity in the disc of intermediate mass star HD 163296, a bright disc around a Herbig star that has been found to have a series of dust-depleted rings \citep{Isella2016}. We have been able to confirm this now for HD 100546, adding further evidence for the prediction of \citet{Wright2015} who suggest grains may have grown to at least 5~cm. \citet{Walsh2014} compare their measured flux with previous mm observations by \citet{Henning1994,Wilner2003} in order to calculate disc-averaged $\beta_{\rm 3mm-1mm}$=0.7-0.8, and $\beta_{\rm 1mm-870\mu m}$=-0.4, values that lie within the range we see over the extent of the disc in Figure \ref{fig:spec_index}. 


We have assumed in our analysis that dust grains in the disc have a size distribution of $dn/da \propto a^{-p}$, where p=3.5, an index typical of the ISM \citep{Mathis1977} that is canonically adopted when modelling discs.
As grains grow, the overall distribution of mass between grain sizes is likely to evolve. \citet{Birnstiel2011DustFits} showed that when both coagulation and fragmentation are included within a disc grain growth model, the size distribution will deviate from p=3.5; growth of grains tends to lower p, whereas fragmentation will re-introduce small particles and raise p. \citet{Draine2006} demonstrates how $\beta$ decreases with index p for a constant $a_{\rm max}$ in Figure 6 of that paper. We would expect ISM-like grains at the outer edge of the disc, yet this is not shown in Figure \ref{fig:spec_index}. Either we lack the sensitivity to detect the full extent of the dust, or the dust in the disc would be better suited to an alternative value of p. The latter would corroborate with the results of \citet{Wright2015} when studying HD 100546, where the authors find that only \citet{dAlessio2005} disc models with p=2.5 fit their data. Tighter constraints could be achieved by comparing against an extensive grid of opacity calculations at millimetre wavelengths over a range of p values.

\section{Conclusions}
\label{sec:conc}
We have presented new ALMA observations of the HD 100546 protoplanetary disc, including the first detection of C$^{18}$O in the disc. 

We measure an integrated flux of 492~mJy from the 1.3~mm continuum image. We derive an outer-most detection of the dust disc at a radial separation from the star of 290$\pm$10~au at 1.3~mm and the lower limit on dust disc mass was calculated to be 4.1-9.6 $\times 10^{-4}~$M$_{\odot}$. 

We detect line emission from CO isotopologue transitions $^{13}$CO (2-1) and C$^{18}$O (2-1).
$^{13}$CO has an integrated flux of 12.872~Jy~kms$^{-1}$ and extends out to 390$\pm$10~au, a size comparable with previous observations of $^{12}$CO in the disc. The optically thinner, mid-plane tracing isotopologue C$^{18}$O is detected for the first time in the disc, giving an integrated flux of 2.948~Jy~kms$^{-1}$ with emission detected as far as 300$\pm$10~au.

From these detections we calculate a considerable gas mass of 0.018 M$_{\odot}$, around 1\% of the stellar mass. We do not expect freeze-out in this warm Herbig disc. If carbon sequestration is efficient the total gas mass may be larger, but not lower, by our calculations. 
The ALMA observations of continuum and CO isotopologue emission presented here suggest the lower limit of the gas to dust ratio for the HD 100546 disc is 19. We have assumed ISM abundances of C$^{18}$O relative to H$_2$ in the disc and so the gas to dust ratio in the disc will be larger in the event that C$^{18}$O is significantly depleted relative to other species.

We observe low-level extended C$^{18}$O emission that is spatially coincident with spiral arm structures identified in scattered light observations by the HST. The extended C$^{18}$O emission may be tracing the mid-plane density variation that creates the scattered light features.
In each of our observed tracers, the disc is seen to have a puzzling offset from the known stellar position of between $\approx$70-110 milliarcseconds.

Spectral index, $\alpha_{\rm mm}$, is derived as a function of radius in the disc and rises with separation from the star suggesting a decrease of maximum dust grain size in the outward direction. We directly relate this to the size of grains through calculating the opacity index, $\beta_{\rm mm}$. This gives results consistent with the growth of grains beyond mm-sizes throughout the disc, and to larger cm-sized `pebbles' within $\sim$200~au. A shallower dust grain size distribution is the alternative explanation, suggesting a larger proportion of the solid mass is contained within the largest grains in comparison to the ISM, or indeed optically thick dust at 867$\mu$m. 

\section*{Acknowledgements}
\small{JM is funded through the University Research Scholarship from the University of Leeds. The work of OP is supported by the Royal Society Dorothy Hodgkin Fellowship. TJH is funded by an Imperial College Junior Research Fellowship. We thank Catherine Walsh for useful advice and discussion. We thank David Ardila for providing the HST images of HD~100546.
This paper makes use of the following ALMA data: ADS/JAO.ALMA$\#$2015.1.01600.S, ADS/JAO.ALMA$\#$2011.0.00863.S . ALMA is a partnership of ESO (representing its member states), NSF (USA) and NINS (Japan), together with NRC (Canada), NSC and ASIAA (Taiwan), and KASI (Republic of Korea), in cooperation with the Republic of Chile. The Joint ALMA Observatory is operated by ESO, AUI/NRAO and NAOJ. 
This research made use of APLpy, an open-source plotting package for Python hosted at http://aplpy.github.com, of Astropy, a community-developed core Python package for Astronomy \citep{TheAstropyCollaboration2013}, of NASA's Astrophysics Data System and the SIMBAD database, operated at CDS, Strasbourg, France.}



\bibliographystyle{mnras}
\bibliography{Mendeley.bib} 

\begin{thebibliography}{}
\makeatletter
\relax
\def\mn@urlcharsother{\let\do\@makeother \do\$\do\&\do\#\do\^\do\_\do\%\do\~}
\def\mn@doi{\begingroup\mn@urlcharsother \@ifnextchar [ {\mn@doi@}
  {\mn@doi@[]}}
\def\mn@doi@[#1]#2{\def\@tempa{#1}\ifx\@tempa\@empty \href
  {http://dx.doi.org/#2} {doi:#2}\else \href {http://dx.doi.org/#2} {#1}\fi
  \endgroup}
\def\mn@eprint#1#2{\mn@eprint@#1:#2::\@nil}
\def\mn@eprint@arXiv#1{\href {http://arxiv.org/abs/#1} {{\tt arXiv:#1}}}
\def\mn@eprint@dblp#1{\href {http://dblp.uni-trier.de/rec/bibtex/#1.xml}
  {dblp:#1}}
\def\mn@eprint@#1:#2:#3:#4\@nil{\def\@tempa {#1}\def\@tempb {#2}\def\@tempc
  {#3}\ifx \@tempc \@empty \let \@tempc \@tempb \let \@tempb \@tempa \fi \ifx
  \@tempb \@empty \def\@tempb {arXiv}\fi \@ifundefined
  {mn@eprint@\@tempb}{\@tempb:\@tempc}{\expandafter \expandafter \csname
  mn@eprint@\@tempb\endcsname \expandafter{\@tempc}}}

\bibitem[\protect\citeauthoryear{Ansdell et~al.,}{Ansdell
  et~al.}{2016}]{Ansdell2016}
Ansdell M.,  et~al., 2016, \mn@doi [ApJ] {10.3847/0004-637X/828/1/46}, 828, 46

\bibitem[\protect\citeauthoryear{Ardila, Golimowski, Krist, Clampin, Ford  \&
  Illingworth}{Ardila et~al.}{2007}]{Ardila2007}
Ardila D.~R.,  Golimowski D.~A.,  Krist J.~E.,  Clampin M.,  Ford H.~C.,
  Illingworth G.~D.,  2007, \mn@doi [ApJ] {10.1086/519296}, 665, 512

\bibitem[\protect\citeauthoryear{Avenhaus, Quanz, Meyer, Brittain, Carr  \&
  Najita}{Avenhaus et~al.}{2014}]{Avenhaus2014}
Avenhaus H.,  Quanz S.~P.,  Meyer M.~R.,  Brittain S.~D.,  Carr J.~S.,   Najita
  J.~R.,  2014, \mn@doi [ApJ] {10.1088/0004-637X/790/1/56}, 790, 56

\bibitem[\protect\citeauthoryear{Banzatti, Testi, Isella, Natta, Neri  \&
  Wilner}{Banzatti et~al.}{2011}]{Banzatti2011AstronomyTauri}
Banzatti A.,  Testi L.,  Isella A.,  Natta A.,  Neri R.,   Wilner D.~J.,  2011,
  \mn@doi [A{\&}A] {10.1051/0004-6361/201015206}, 525

\bibitem[\protect\citeauthoryear{Baruteau, Crida, Paardekooper, Guilet, Bitsch,
  Nelson, Kley  \& Papaloizou}{Baruteau et~al.}{2014}]{Baruteau2014}
Baruteau C.,  Crida A.,  Paardekooper S.-J.,  Guilet J.,  Bitsch B.,  Nelson
  R.,  Kley W.,   Papaloizou J.,  2014, in {Henrik Beuther} {Ralf S. Klessen}
  {Cornelis P. Dullemond}  {Thomas Henning} eds, , PPVI.
Univ.of Arizona Press, Tucson, pp 667--689, \url
  {https://arxiv.org/pdf/1312.4293.pdf}

\bibitem[\protect\citeauthoryear{Birnstiel \& Andrews}{Birnstiel \&
  Andrews}{2013}]{Birnstiel2013}
Birnstiel T.,  Andrews S.~M.,  2013, \mn@doi [ApJ]
  {10.1088/0004-637X/780/2/153}, 780, 153

\bibitem[\protect\citeauthoryear{Birnstiel et~al.,}{Birnstiel
  et~al.}{2010}]{Birnstiel2010b}
Birnstiel T.,  et~al., 2010, \mn@doi [A{\&}A] {10.1051/0004-6361/201014893},
  516, L14

\bibitem[\protect\citeauthoryear{Birnstiel, Ormel  \& Dullemond}{Birnstiel
  et~al.}{2011}]{Birnstiel2011DustFits}
Birnstiel T.,  Ormel C.~W.,   Dullemond C.~P.,  2011, \mn@doi [A{\&}A]
  {10.1051/0004-6361/201015228}, 525, A11

\bibitem[\protect\citeauthoryear{Boneberg, Pani{\'{c}}, Haworth, Clarke  \&
  Min}{Boneberg et~al.}{2016}]{Boneberg2016}
Boneberg D.~M.,  Pani{\'{c}} O.,  Haworth T.~J.,  Clarke C.~J.,   Min M.,
  2016, \mn@doi [MNRAS] {10.1093/mnras/stw1325}, 461, 385

\bibitem[\protect\citeauthoryear{Brauer, Dullemond  \& Henning}{Brauer
  et~al.}{2008}]{Brauer2008}
Brauer F.,  Dullemond C.~P.,   Henning T.,  2008, \mn@doi [A{\&}A]
  {10.1051/0004-6361:20077759}, 480, 859

\bibitem[\protect\citeauthoryear{Bressan, Marigo, Girardi, Salasnich, Cero,
  Rubele  \& Nanni}{Bressan et~al.}{2012}]{Bressan2012}
Bressan A.,  Marigo P.,  Girardi L.,  Salasnich B.,  Cero C.~D.,  Rubele S.,
  Nanni A.,  2012, \mn@doi [MNRAS] {10.1111/j.1365-2966.2012.21948.x}, 427, 127

\bibitem[\protect\citeauthoryear{Brittain, Najita, Carr, Liskowsky, Troutman
  \& Doppmann}{Brittain et~al.}{2013}]{Brittain2013}
Brittain S.~D.,  Najita J.~R.,  Carr J.~S.,  Liskowsky J.,  Troutman M.~R.,
  Doppmann G.~W.,  2013, \mn@doi [ApJ] {10.1088/0004-637X/767/2/159}, 767, 159

\bibitem[\protect\citeauthoryear{Brittain, Carr, Najita, Quanz  \&
  Meyer}{Brittain et~al.}{2014}]{Brittain2014}
Brittain S.~D.,  Carr J.~S.,  Najita J.~R.,  Quanz S.~P.,   Meyer M.~R.,  2014,
  \mn@doi [ApJ] {10.1088/0004-637X/791/2/136}, 791, 136

\bibitem[\protect\citeauthoryear{Bruderer, van Dishoeck, Doty  \&
  Herczeg}{Bruderer et~al.}{2012}]{Bruderer2012}
Bruderer S.,  van Dishoeck E.~F.,  Doty S.~D.,   Herczeg G.~J.,  2012, \mn@doi
  [A{\&}A] {10.1051/0004-6361/201118218}, 541, A91

\bibitem[\protect\citeauthoryear{Casassus et~al.,}{Casassus
  et~al.}{2015}]{Casassus2015}
Casassus S.,  et~al., 2015, \mn@doi [ApJ] {10.1088/0004-637X/812/2/126}, 812,
  126

\bibitem[\protect\citeauthoryear{Cazzoletti et~al.,}{Cazzoletti
  et~al.}{2018}]{Cazzoletti2018}
Cazzoletti P.,  et~al., 2018, \mn@doi [A{\&}A] {10.1051/0004-6361/201834006},
  619, A161

\bibitem[\protect\citeauthoryear{Currie et~al.,}{Currie
  et~al.}{2015}]{Currie2015}
Currie T.,  et~al., 2015, \mn@doi [ApJ] {10.1088/2041-8205/814/2/L27}, 814, L27

\bibitem[\protect\citeauthoryear{Dartois, Dutrey  \& Guilloteau}{Dartois
  et~al.}{2003}]{Dartois2003}
Dartois E.,  Dutrey A.,   Guilloteau S.,  2003, \mn@doi [A{\&}A]
  {10.1051/0004-6361:20021638}, 399, 773

\bibitem[\protect\citeauthoryear{Draine}{Draine}{2006}]{Draine2006}
Draine B.~T.,  2006, \mn@doi [ApJ] {10.1086/498130}, 636, 1114

\bibitem[\protect\citeauthoryear{Du, Bergin  \& Hogerheijde}{Du
  et~al.}{2015}]{Du2015}
Du F.,  Bergin E.~A.,   Hogerheijde M.~R.,  2015, \mn@doi [ApJL]
  {10.1088/2041-8205/807/2/L32}, 807, L32

\bibitem[\protect\citeauthoryear{Dullemond et~al.,}{Dullemond
  et~al.}{2018}]{Dullemond2018}
Dullemond C.~P.,  et~al., 2018, \mn@doi [ApJL] {10.3847/2041-8213/aaf742}, 869,
  L46

\bibitem[\protect\citeauthoryear{Dutrey et~al.,}{Dutrey
  et~al.}{2014}]{Dutrey2014PossibleA}
Dutrey A.,  et~al., 2014, \mn@doi [Nature] {10.1038/nature13822}, 514, 600

\bibitem[\protect\citeauthoryear{Dzyurkevich, Flock, Turner, Klahr  \&
  Henning}{Dzyurkevich et~al.}{2010}]{Dzyurkevich2010TrappingSimulations}
Dzyurkevich N.,  Flock M.,  Turner N.~J.,  Klahr H.,   Henning T.,  2010,
  \mn@doi [A{\&}A] {10.1051/0004-6361/200912834}, 515, A70

\bibitem[\protect\citeauthoryear{Ercolano \& Pascucci}{Ercolano \&
  Pascucci}{2017}]{Ercolano2017TheObservations}
Ercolano B.,  Pascucci I.,  2017, \mn@doi [Royal Society Open Science]
  {10.1098/rsos.170114}, 4, 170114

\bibitem[\protect\citeauthoryear{Espaillat et~al.,}{Espaillat
  et~al.}{2014}]{Espaillat2014}
Espaillat C.,  et~al., 2014, in Beuther H.,  Klessen R.~S.,  Dullemond C.~P.,
  Henning T.,  eds, , PPIV.
Univ. Arizona Press, Tucson, p.~475

\bibitem[\protect\citeauthoryear{Facchini, Birnstiel, Bruderer  \&
  Van~Dishoeck}{Facchini et~al.}{2017}]{Facchini2017}
Facchini S.,  Birnstiel T.,  Bruderer S.,   Van~Dishoeck E.~F.,  2017, \mn@doi
  [A{\&}A] {10.1051/0004-6361/201630329}, 605, 16

\bibitem[\protect\citeauthoryear{Fairlamb, Oudmaijer, Mendigut{\'{i}}a, Ilee
  \& van~den Ancker}{Fairlamb et~al.}{2015}]{Fairlamb2015}
Fairlamb J.~R.,  Oudmaijer R.~D.,  Mendigut{\'{i}}a I.,  Ilee J.~D.,   van~den
  Ancker M.~E.,  2015, \mn@doi [MNRAS] {10.1093/mnras/stv1576}, 453, 976

\bibitem[\protect\citeauthoryear{Fedele et~al.,}{Fedele
  et~al.}{2017}]{Fedele2017}
Fedele D.,  et~al., 2017, \mn@doi [A{\&}A] {10.1051/0004-6361/201629860}, 600,
  A72

\bibitem[\protect\citeauthoryear{Follette et~al.,}{Follette
  et~al.}{2017}]{Follette2017}
Follette K.~B.,  et~al., 2017, \mn@doi [AJ] {10.3847/1538-3881/aa6d85}, 153,
  264

\bibitem[\protect\citeauthoryear{Garufi et~al.,}{Garufi
  et~al.}{2016}]{Garufi2016}
Garufi A.,  et~al., 2016, \mn@doi [A{\&}A] {10.1051/0004-6361/201527940}, 588,
  A8

\bibitem[\protect\citeauthoryear{Grady et~al.,}{Grady et~al.}{2001}]{Grady2001}
Grady C.~A.,  et~al., 2001, \mn@doi [AJ] {10.1086/324447}, 122, 3396

\bibitem[\protect\citeauthoryear{Guidi et~al.,}{Guidi et~al.}{2016}]{Guidi2016}
Guidi G.,  et~al., 2016, \mn@doi [A{\&}A] {10.1051/0004-6361/201527516}, 588,
  A112

\bibitem[\protect\citeauthoryear{Guilloteau, Di~Folco, Dutrey, Simon, Grosso
  \& Pi{\'{e}}tu}{Guilloteau et~al.}{2013}]{Guilloteau2013}
Guilloteau S.,  Di~Folco E.,  Dutrey A.,  Simon M.,  Grosso N.,   Pi{\'{e}}tu
  V.,  2013, \mn@doi [A{\&}A] {10.1051/0004-6361/201220298}, 549, A92

\bibitem[\protect\citeauthoryear{Henning, Launhardt, Steinacker  \&
  Thamm}{Henning et~al.}{1994}]{Henning1994}
Henning T.,  Launhardt R.,  Steinacker J.,   Thamm E.,  1994, A{\&}A, 291, 546

\bibitem[\protect\citeauthoryear{Hughes, Wilner, Kamp  \& Hogerheijde}{Hughes
  et~al.}{2008}]{Hughes2008}
Hughes A.~M.,  Wilner D.~J.,  Kamp I.,   Hogerheijde M.~R.,  2008, \mn@doi
  [ApJ] {10.1086/588520}, 681, 626

\bibitem[\protect\citeauthoryear{Isella et~al.,}{Isella
  et~al.}{2016}]{Isella2016}
Isella A.,  et~al., 2016, \mn@doi [PRL] {10.1103/PhysRevLett.117.251101}, 117,
  251101

\bibitem[\protect\citeauthoryear{Juh{\'{a}}sz \& Rosotti}{Juh{\'{a}}sz \&
  Rosotti}{2018}]{Juhasz2018}
Juh{\'{a}}sz A.,  Rosotti G.~P.,  2018, \mn@doi [MNRASL]
  {10.1093/mnrasl/slx182}, 474, L32

\bibitem[\protect\citeauthoryear{Kama et~al.,}{Kama et~al.}{2016}]{Kama2016}
Kama M.,  et~al., 2016, \mn@doi [A{\&}A] {10.1051/0004-6361/201526991}, 592,
  A83

\bibitem[\protect\citeauthoryear{Klahr \& Bodenheimer}{Klahr \&
  Bodenheimer}{2006}]{Klahr2006}
Klahr H.,  Bodenheimer P.,  2006, \mn@doi [ApJ] {10.1086/498928}, 639, 432

\bibitem[\protect\citeauthoryear{Lindegren et~al.,}{Lindegren
  et~al.}{2018}]{Lindegren2018}
Lindegren L.,  et~al., 2018, \mn@doi [A{\&}A] {10.1051/0004-6361/201832727},
  616, A2

\bibitem[\protect\citeauthoryear{Long et~al.,}{Long et~al.}{2017}]{Long2017}
Long F.,  et~al., 2017, \mn@doi [ApJ] {10.3847/1538-4357/aa78fc}, 844, 99

\bibitem[\protect\citeauthoryear{Loomis, {\"{O}}berg, Andrews  \&
  MacGregor}{Loomis et~al.}{2017}]{Loomis2017}
Loomis R.~A.,  {\"{O}}berg K.~I.,  Andrews S.~M.,   MacGregor M.~A.,  2017,
  \mn@doi [ApJ] {10.3847/1538-4357/aa6c63}, 840, 23

\bibitem[\protect\citeauthoryear{Manara, Morbidelli  \& Guillot}{Manara
  et~al.}{2018}]{Manara2018WhyPopulation}
Manara C.~F.,  Morbidelli A.,   Guillot T.,  2018, \mn@doi [A{\&}A]
  {10.1051/0004-6361/201834076}, 618

\bibitem[\protect\citeauthoryear{Mathis, Rumpl  \& Nordsieck}{Mathis
  et~al.}{1977}]{Mathis1977}
Mathis J.~S.,  Rumpl W.,   Nordsieck K.~H.,  1977, ApJ, 217, 425

\bibitem[\protect\citeauthoryear{Matr{\`{a}}, Pani{\'{c}}, Wyatt  \&
  Dent}{Matr{\`{a}} et~al.}{2015}]{Matra2015}
Matr{\`{a}} L.,  Pani{\'{c}} O.,  Wyatt M.~C.,   Dent W. R.~F.,  2015, \mn@doi
  [MNRAS] {10.1093/mnras/stu2619}, 447, 3936

\bibitem[\protect\citeauthoryear{Meeus et~al.,}{Meeus et~al.}{2010}]{Meeus2010}
Meeus G.,  et~al., 2010, \mn@doi [A{\&}A] {10.1051/0004-6361/201014557}, 518

\bibitem[\protect\citeauthoryear{Mendigut{\'{i}}a et~al.,}{Mendigut{\'{i}}a
  et~al.}{2017}]{Mendigutia2017b}
Mendigut{\'{i}}a I.,  et~al., 2017, \mn@doi [A{\&}A]
  {10.1051/0004-6361/201731131}, 608, A104

\bibitem[\protect\citeauthoryear{Miley, Panic, Wyatt  \& Kennedy}{Miley
  et~al.}{2018}]{Miley2018}
Miley J.,  Panic O.,  Wyatt M.,   Kennedy G.,  2018, \mn@doi [A{\&}A]
  {10.1051/0004-6361/201833381}, 615, L10

\bibitem[\protect\citeauthoryear{Miotello, Bruderer  \& van Dishoeck}{Miotello
  et~al.}{2014}]{Miotello2014}
Miotello A.,  Bruderer S.,   van Dishoeck E.~F.,  2014, \mn@doi [A{\&}A]
  {10.1051/0004-6361/201424712}, 572, A96

\bibitem[\protect\citeauthoryear{Miotello, van Dishoeck, Kama  \&
  Bruderer}{Miotello et~al.}{2016}]{Miotello2016}
Miotello A.,  van Dishoeck E.~F.,  Kama M.,   Bruderer S.,  2016, \mn@doi
  [A{\&}A] {10.1051/0004-6361/201628159}, 594, A85

\bibitem[\protect\citeauthoryear{Miotello et~al.,}{Miotello
  et~al.}{2017}]{Miotello2017}
Miotello A.,  et~al., 2017, \mn@doi [A{\&}A] {10.1051/0004-6361/201629556},
  599, A113

\bibitem[\protect\citeauthoryear{Mo{\'{o}}r et~al.,}{Mo{\'{o}}r
  et~al.}{2011}]{Moor2011}
Mo{\'{o}}r A.,  et~al., 2011, \mn@doi [ApJ] {10.1088/2041-8205/740/1/L7}, 740,
  L7

\bibitem[\protect\citeauthoryear{Mordasini}{Mordasini}{2018}]{Mordasini2018}
Mordasini C.,  2018, in Deeg H.~J.,  Belmonte J.~A.,  eds, , Handbook of
  Exoplanets.
Springer, Cham, \mn@doi{10.1007/978-3-319-30648-3{\_}143-1}, \url
  {https://arxiv.org/pdf/1804.01532.pdf http://arxiv.org/abs/1804.01532}

\bibitem[\protect\citeauthoryear{Mordasini, Alibert, Benz, Klahr  \&
  Henning}{Mordasini et~al.}{2012}]{Mordasini2012AstrophysicsLifetime}
Mordasini C.,  Alibert Y.,  Benz W.,  Klahr H.,   Henning T.,  2012, \mn@doi
  [A{\&}A] {10.1051/0004-6361/201117350}, 541, A97

\bibitem[\protect\citeauthoryear{Mulders, Paardekooper, Pani{\'{c}}, Dominik,
  van Boekel  \& Ratzka}{Mulders et~al.}{2013}]{Mulders2013}
Mulders G.~D.,  Paardekooper S.-J.,  Pani{\'{c}} O.,  Dominik C.,  van Boekel
  R.,   Ratzka T.,  2013, \mn@doi [A{\&}A] {10.1051/0004-6361/201220930}, 557,
  A68

\bibitem[\protect\citeauthoryear{Owen}{Owen}{2016}]{Owen2016}
Owen J.~E.,  2016, \mn@doi [Publications of the Astronomical Society of
  Australia] {10.1017/pasa.2016.2}, 33, e005

\bibitem[\protect\citeauthoryear{Pacheco-Vazquez et~al.,}{Pacheco-Vazquez
  et~al.}{2016}]{Pacheco-Vazquez2016}
Pacheco-Vazquez S.,  et~al., 2016, \mn@doi [A{\&}A]
  {10.1051/0004-6361/201527089}, 589, A60

\bibitem[\protect\citeauthoryear{Pani{\'{c}} \& Hogerheijde}{Pani{\'{c}} \&
  Hogerheijde}{2009}]{Panic2009CharacterisingLines}
Pani{\'{c}} O.,  Hogerheijde M.~R.,  2009, \mn@doi [A{\&}A]
  {10.1051/0004-6361/200912584}, 508, 707

\bibitem[\protect\citeauthoryear{Pani{\'{c}} \& Min}{Pani{\'{c}} \&
  Min}{2017}]{Panic2017}
Pani{\'{c}} O.,  Min M.,  2017, \mn@doi [MNRAS] {10.1093/mnras/stx114}, 467,
  1175

\bibitem[\protect\citeauthoryear{Pani{\'{c}}, Hogerheijde, Wilner  \&
  Qi}{Pani{\'{c}} et~al.}{2008}]{Panic2008}
Pani{\'{c}} O.,  Hogerheijde M.~R.,  Wilner D.,   Qi C.,  2008, \mn@doi
  [A{\&}A] {10.1051/0004-6361:20079261}, 491, 219

\bibitem[\protect\citeauthoryear{Pani{\'{c}}, van Dishoeck, Hogerheijde,
  Belloche, G{\"{u}}sten, Boland  \& Baryshev}{Pani{\'{c}}
  et~al.}{2010}]{Panic2010Warm100546}
Pani{\'{c}} O.,  van Dishoeck E.~F.,  Hogerheijde M.~R.,  Belloche A.,
  G{\"{u}}sten R.,  Boland W.,   Baryshev A.,  2010, \mn@doi [A{\&}A]
  {10.1051/0004-6361/200913709}, 519

\bibitem[\protect\citeauthoryear{Pani{\'{c}}, Ratzka, Mulders, Dominik, van
  Boekel, Henning, Jaffe  \& Min}{Pani{\'{c}} et~al.}{2014}]{Panic2014}
Pani{\'{c}} O.,  Ratzka T.,  Mulders G.~D.,  Dominik C.,  van Boekel R.,
  Henning T.,  Jaffe W.,   Min M.,  2014, \mn@doi [A{\&}A]
  {10.1051/0004-6361/201219223}, 562, A101

\bibitem[\protect\citeauthoryear{P{\'{e}}rez et~al.,}{P{\'{e}}rez
  et~al.}{2012}]{Perez2012CONSTRAINTSDISK}
P{\'{e}}rez L.~M.,  et~al., 2012, \mn@doi [ApJ] {10.1088/2041-8205/760/1/L17},
  760, L17

\bibitem[\protect\citeauthoryear{Perez et~al.,}{Perez et~al.}{2014}]{Perez2014}
Perez S.,  et~al., 2014, \mn@doi [ApJ] {10.1088/0004-637X/798/2/85}, 798, 85

\bibitem[\protect\citeauthoryear{Pineda, Quanz, Meru, Mulders, Meyer,
  Pani{\'{c}}  \& Avenhaus}{Pineda et~al.}{2014}]{Pineda2014}
Pineda J.~E.,  Quanz S.~P.,  Meru F.,  Mulders G.~D.,  Meyer M.~R.,
  Pani{\'{c}} O.,   Avenhaus H.,  2014, \mn@doi [ApJL]
  {10.1088/2041-8205/788/2/L34}, 788

\bibitem[\protect\citeauthoryear{Pineda et~al.,}{Pineda
  et~al.}{2018}]{Pineda2018High-ResolutionLimits}
Pineda J.~E.,  et~al., 2018, ApJ

\bibitem[\protect\citeauthoryear{Pinilla, Birnstiel, Ricci, Dullemond, Uribe,
  Testi  \& Natta}{Pinilla et~al.}{2012a}]{Pinilla2012}
Pinilla P.,  Birnstiel T.,  Ricci L.,  Dullemond C.~P.,  Uribe A.~L.,  Testi
  L.,   Natta A.,  2012a, \mn@doi [A{\&}A] {10.1051/0004-6361/201118204}, 538,
  A114

\bibitem[\protect\citeauthoryear{Pinilla, Benisty  \& Birnstiel}{Pinilla
  et~al.}{2012b}]{Pinilla2012RingDisks}
Pinilla P.,  Benisty M.,   Birnstiel T.,  2012b, \mn@doi [A{\&}A]
  {10.1051/0004-6361/201219315}, 545, A81

\bibitem[\protect\citeauthoryear{Pinilla et~al.,}{Pinilla
  et~al.}{2014}]{Pinilla2014}
Pinilla P.,  et~al., 2014, \mn@doi [A{\&}A] {10.1051/0004-6361/201323322}, 564,
  A51

\bibitem[\protect\citeauthoryear{Pinilla et~al.,}{Pinilla
  et~al.}{2018}]{Pinilla2018HomogeneousCavities}
Pinilla P.,  et~al., 2018, \mn@doi [ApJ] {10.3847/1538-4357/aabf94}, 859, 32

\bibitem[\protect\citeauthoryear{Price et~al.,}{Price et~al.}{2018}]{Price2018}
Price D.~J.,  et~al., 2018, \mn@doi [MNRAS] {10.1093/mnras/sty647}, 477, 1270

\bibitem[\protect\citeauthoryear{Quanz, Amara, Meyer, Girard, Kenworthy  \&
  Kasper}{Quanz et~al.}{2015}]{Quanz2015CONFIRMATIONAU}
Quanz S.~P.,  Amara A.,  Meyer M.~R.,  Girard J.~H.,  Kenworthy M.~A.,   Kasper
  M.,  2015, \mn@doi [ApJ] {10.1088/0004-637X/807/1/64}, 807, 64

\bibitem[\protect\citeauthoryear{Quillen}{Quillen}{2006}]{Quillen2006}
Quillen A.~C.,  2006, \mn@doi [ApJ] {10.1086/500165}, 640, 1078

\bibitem[\protect\citeauthoryear{Rau \& Cornwell}{Rau \&
  Cornwell}{2011}]{Rau2011}
Rau U.,  Cornwell T.~J.,  2011, \mn@doi [A{\&}A] {10.1051/0004-6361/201117104},
  532, A71

\bibitem[\protect\citeauthoryear{Reboussin, Wakelam, Guilloteau, Hersant  \&
  Dutrey}{Reboussin et~al.}{2015}]{Reboussin2015}
Reboussin L.,  Wakelam V.,  Guilloteau S.,  Hersant F.,   Dutrey A.,  2015,
  \mn@doi [A{\&}A] {10.1051/0004-6361/201525885}, 579, A82

\bibitem[\protect\citeauthoryear{Ricci, Trotta, Testi, Natta, Isella  \&
  Wilner}{Ricci et~al.}{2012}]{Ricci2012}
Ricci L.,  Trotta F.,  Testi L.,  Natta A.,  Isella A.,   Wilner D.~J.,  2012,
  \mn@doi [A{\&}A] {10.1051/0004-6361/201118296}, 540, A6

\bibitem[\protect\citeauthoryear{Rosenfeld, Andrews, Hughes, Wilner  \&
  Qi}{Rosenfeld et~al.}{2013}]{Rosenfeld2013}
Rosenfeld K.~A.,  Andrews S.~M.,  Hughes A.~M.,  Wilner D.~J.,   Qi C.,  2013,
  \mn@doi [ApJ] {10.1088/0004-637X/774/1/16}, 774, 16

\bibitem[\protect\citeauthoryear{Schwarz, Bergin, Cleeves, Blake, Zhang,
  {\"{O}}berg, van Dishoeck  \& Qi}{Schwarz et~al.}{2016}]{Schwarz2016}
Schwarz K.~R.,  Bergin E.~A.,  Cleeves L.~I.,  Blake G.~A.,  Zhang K.,
  {\"{O}}berg K.~I.,  van Dishoeck E.~F.,   Qi C.,  2016, \mn@doi [ApJ]
  {10.3847/0004-637X/823/2/91}, 823, 91

\bibitem[\protect\citeauthoryear{Takeuchi \& Lin}{Takeuchi \&
  Lin}{2005}]{Takeuchi2005}
Takeuchi T.,  Lin D. N.~C.,  2005, \mn@doi [ApJ] {10.1086/428378}, 623, 482

\bibitem[\protect\citeauthoryear{Tanaka, Himeno  \& Ida}{Tanaka
  et~al.}{2005}]{Tanaka2005}
Tanaka H.,  Himeno Y.,   Ida S.,  2005, \mn@doi [ApJ] {10.1086/429658}, 625,
  414

\bibitem[\protect\citeauthoryear{Tazzari et~al.,}{Tazzari
  et~al.}{2016}]{Tazzari2016}
Tazzari M.,  et~al., 2016, \mn@doi [A{\&}A] {10.1051/0004-6361/201527423}, 588,
  A53

\bibitem[\protect\citeauthoryear{{The Astropy Collaboration} et~al.,}{{The
  Astropy Collaboration} et~al.}{2013}]{TheAstropyCollaboration2013}
{The Astropy Collaboration} et~al., 2013, \mn@doi [A{\&}A]
  {10.1051/0004-6361/201322068}, 558, A33

\bibitem[\protect\citeauthoryear{Toomre}{Toomre}{1964}]{Toomre1964}
Toomre A.,  1964, ApJ, 139, 1217

\bibitem[\protect\citeauthoryear{Trotta, Testi, Natta, Isella  \& Ricci}{Trotta
  et~al.}{2013}]{Trotta2013}
Trotta F.,  Testi L.,  Natta A.,  Isella A.,   Ricci L.,  2013, \mn@doi
  [A{\&}A] {10.1051/0004-6361/201321896}, 558, A64

\bibitem[\protect\citeauthoryear{Van Der~Marel, Cazzoletti, Pinilla  \&
  Garufi}{Van Der~Marel et~al.}{2016}]{VanDerMarel2016}
Van Der~Marel N.,  Cazzoletti P.,  Pinilla P.,   Garufi A.,  2016, \mn@doi
  [ApJ] {10.3847/0004-637X/832/2/178}, 823, 178

\bibitem[\protect\citeauthoryear{Van~Zadelhoff, Van~Dishoeck, Thi  \&
  Blake}{Van~Zadelhoff et~al.}{2001}]{VanZadelhoff2001}
Van~Zadelhoff G.-J.,  Van~Dishoeck E.~F.,  Thi W.-F.,   Blake G.~A.,  2001,
  \mn@doi [A{\&}A] {10.1051/0004-6361:20011137}, 377, 566

\bibitem[\protect\citeauthoryear{Vioque, Oudmaijer, Baines, Mendigut{\'{i}}a
  \& P{\'{e}}rez-Mart{\'{i}}nez}{Vioque et~al.}{2018}]{Vioque2018}
Vioque M.,  Oudmaijer R.~D.,  Baines D.,  Mendigut{\'{i}}a I.,
  P{\'{e}}rez-Mart{\'{i}}nez R.,  2018, \mn@doi [A{\&}A]
  {10.1051/0004-6361/201832870}, 620, 128

\bibitem[\protect\citeauthoryear{Walsh et~al.,}{Walsh et~al.}{2014}]{Walsh2014}
Walsh C.,  et~al., 2014, \mn@doi [ApJL] {10.1088/2041-8205/791/1/L6}, 791, L6

\bibitem[\protect\citeauthoryear{Walsh, Daley, Facchini  \& Juh{\'{a}}sz}{Walsh
  et~al.}{2017}]{Walsh2017}
Walsh C.,  Daley C.,  Facchini S.,   Juh{\'{a}}sz A.,  2017, \mn@doi [A{\&}A]
  {10.1051/0004-6361/201731334}, 607, A114

\bibitem[\protect\citeauthoryear{White, Boley, Hughes, Flaherty, Ford, Wilner,
  Corder  \& Payne}{White et~al.}{2016}]{White2016}
White J.~A.,  Boley A.~C.,  Hughes A.~M.,  Flaherty K.~M.,  Ford E.,  Wilner
  D.,  Corder S.,   Payne M.,  2016, \mn@doi [ApJ] {10.3847/0004-637X/829/1/6},
  829, 6

\bibitem[\protect\citeauthoryear{Williams \& Cieza}{Williams \&
  Cieza}{2011}]{Williams2011}
Williams J.~P.,  Cieza L.~A.,  2011, \mn@doi [ARA{\&}A]
  {10.1146/annurev-astro-081710-102548}, 49, 67

\bibitem[\protect\citeauthoryear{Williams et~al.,}{Williams
  et~al.}{2014}]{Williams2014}
Williams J.~P.,  et~al., 2014, \mn@doi [ApJ] {10.1088/0004-637X/796/2/120},
  796, 120

\bibitem[\protect\citeauthoryear{Wilner, Bourke, Wright, J{\o}rgensen,
  Van~Dishoeck  \& Wong}{Wilner et~al.}{2003}]{Wilner2003}
Wilner D.~J.,  Bourke T.~L.,  Wright C.~M.,  J{\o}rgensen J.~K.,  Van~Dishoeck
  E.~F.,   Wong T.,  2003, ApJ, 596, 597

\bibitem[\protect\citeauthoryear{Wilson}{Wilson}{1999}]{Wilson1999}
Wilson T.~L.,  1999, \mn@doi [RPPh] {10.1088/0034-4885/62/2/002}, 62, 143

\bibitem[\protect\citeauthoryear{Wilson \& Rood}{Wilson \&
  Rood}{1994}]{Wilson1994}
Wilson T.~L.,  Rood R.~T.,  1994, \mn@doi [ARA{\&}A]
  {10.1146/annurev.aa.32.090194.001203}, 32, 191

\bibitem[\protect\citeauthoryear{Wright et~al.,}{Wright
  et~al.}{2015}]{Wright2015}
Wright C.~M.,  et~al., 2015, \mn@doi [MNRAS] {10.1093/mnras/stv1619}, 453, 414

\bibitem[\protect\citeauthoryear{Wyatt, Pani{\'{c}}, Kennedy  \&
  Matr{\`{a}}}{Wyatt et~al.}{2015}]{Wyatt2015}
Wyatt M.~C.,  Pani{\'{c}} O.,  Kennedy G.~M.,   Matr{\`{a}} L.,  2015, \mn@doi
  [Ap{\&}SS] {10.1007/s10509-015-2315-6}, 357, 103

\bibitem[\protect\citeauthoryear{Yu, Willacy, Dodson-Robinson, Turner  \&
  Evans~Ii}{Yu et~al.}{2016}]{Yu2016}
Yu M.,  Willacy K.,  Dodson-Robinson S.~E.,  Turner N.~J.,   Evans~Ii N.~J.,
  2016, \mn@doi [ApJ] {10.3847/0004-637X/822/1/53}, 822, 53

\bibitem[\protect\citeauthoryear{d 'Alessio, Mer{\'{i}}n, Calvet, Hartmann  \&
  Montesinos}{d~'Alessio et~al.}{2005}]{dAlessio2005}
d 'Alessio P.,  Mer{\'{i}}n B.,  Calvet N.,  Hartmann L.,   Montesinos B.,
  2005, RMxAA, 41, 61

\bibitem[\protect\citeauthoryear{van~den Ancker, Th{\'{e}}, Tjin, Djie, Catala,
  De~Winter, Blondel  \& Waters}{van~den Ancker
  et~al.}{1997}]{vandenAncker1997}
van~den Ancker M.~E.,  Th{\'{e}} P.~S.,  Tjin H. R.~E.,  Djie A.,  Catala C.,
  De~Winter D.,  Blondel P. F.~C.,   Waters L. B. F.~M.,  1997, A{\&}A, 324, 33

\bibitem[\protect\citeauthoryear{van~der Marel}{van~der
  Marel}{2017}]{vanderMarel2017}
van~der Marel N.,  2017, in Pessah M.,  Gressel O.,  eds, , Formation,
  Evolution, and Dynamics of Young Solar Systems.
Springer, Cham, pp 39--61, \mn@doi{10.1007/978-3-319-60609-5{\_}2}, \url
  {http://link.springer.com/10.1007/978-3-319-60609-5_2}

\bibitem[\protect\citeauthoryear{van~der Marel et~al.,}{van~der Marel
  et~al.}{2013}]{vanderMarel2013ADisk}
van~der Marel N.,  et~al., 2013, \mn@doi [Science] {10.1126/science.1236770},
  340, 1199

\bibitem[\protect\citeauthoryear{van~der Marel et~al.,}{van~der Marel
  et~al.}{2015}]{vanderMarel2015}
van~der Marel N.,  et~al., 2015, \mn@doi [ApJ] {10.1088/2041-8205/810/1/L7},
  810, L7

\makeatother
\end{thebibliography}



\bsp	
\label{lastpage}
\end{document}